\documentclass[a4paper,11pt]{article}

\usepackage[margin=1in]{geometry}

\usepackage[english]{babel}
\usepackage{amssymb}
\usepackage{amsmath}
\usepackage{amsthm}
\usepackage{psfrag}
\usepackage[T1]{fontenc}
\usepackage{ae,aecompl}
\usepackage[colorlinks]{hyperref}
\usepackage{subfigure}
\usepackage{appendix}
\usepackage{natbib}
\hypersetup{colorlinks,breaklinks=true,citecolor=blue,linkcolor=blue}
\usepackage{graphicx}
\usepackage{color}

\providecommand\bnabla{\boldsymbol{\nabla}}
\providecommand\bcdot{\boldsymbol{\cdot}}
\newcommand\ie{i.e.\ }
\newcommand\eg{e.g.\ }

\newcommand{\pleft}{\left(}
\newcommand{\pright}{\right)}
\newcommand{\vect}[1]{\boldsymbol{#1}}
\newcommand{\B}{\boldsymbol{B}}
\newcommand{\vel}{\boldsymbol{u}}

\newcommand{\pdt}[1]{\frac{\partial #1}{\partial t}}
\newcommand{\Ek}{\mbox{\textit{Ek}}}
\newcommand{\Ra}{\mbox{\textit{Ra}}}

\newcommand{\Pm}{\mbox{\textit{Pm}}}
\newcommand{\Ro}{\mbox{\textit{Ro}}}

\newcommand{\Rm}{\mbox{\textit{Rm}}}
\newcommand{\Rey}{\mbox{\textit{Re}}}
\newcommand{\Pran}{\mbox{\textit{Pr}}}
\newcommand{\aspect}{\lambda}

\title{Large-scale-vortex dynamos in planar rotating convection}

\author{C\'eline Guervilly\footnote{Present address: School of Mathematics and Statistics, Newcastle University, Newcastle Upon Tyne, NE17RU UK.}, David W. Hughes \& Chris A. Jones \vspace{0.2cm}
 \\ {\small Department of Applied Mathematics, University of Leeds, Leeds LS2 9JT, UK}}

\begin{document}

\maketitle

\begin{abstract}
Several recent studies have demonstrated how large-scale vortices may arise spontaneously in rotating planar convection. Here we examine the dynamo properties of such flows in rotating Boussinesq convection. For moderate values of the magnetic Reynolds number ($100 \lesssim \Rm \lesssim 550$, with $\Rm$ based on the box depth and the convective velocity), a large-scale (\ie system-size) magnetic field is generated. The amplitude of the magnetic energy oscillates in time, nearly out of phase with the oscillating amplitude of the large-scale vortex. 
The large-scale vortex is disrupted once the magnetic field reaches a critical strength, showing that these oscillations are of magnetic origin.
The dynamo mechanism relies on those components of the flow that have length scales lying between that of the large-scale vortex and the typical convective cell size; smaller-scale flows are not required. The large-scale vortex plays a crucial role in the magnetic induction despite being essentially two-dimensional;
we thus refer to this dynamo as a large-scale-vortex dynamo. 
For larger magnetic Reynolds numbers, the dynamo is small scale, with a magnetic energy spectrum that peaks at the scale of the convective cells. In this case, the small-scale magnetic field 
continuously suppresses the large-scale vortex by disrupting the correlations between the convective velocities that allow it to form. 
The suppression of the large-scale vortex at high $\Rm$ therefore probably limits the relevance of the large-scale-vortex dynamo 
to astrophysical objects with moderate values of $\Rm$, such as planets. In this context, the ability of the large-scale-vortex dynamo to operate at low magnetic Prandtl numbers 
is of great interest.
\end{abstract}

%\begin{keywords}
%rotating turbulence, turbulent convection, dynamo theory
%\end{keywords}

\section{Introduction}

Understanding the generation of system-size magnetic fields in natural objects, \ie fields with a significant component on the scale of the objects themselves, remains an outstanding problem in geophysical and astrophysical fluid dynamics. Such fields are maintained by dynamo action, whereby the magnetic induction produced by the motions of an electrically conducting fluid compensates the losses due to Ohmic dissipation. Typically, in planetary and stellar interiors, the inductive motions are driven by thermal 
or compositional convection.

Numerical simulations have demonstrated that rotating convection can indeed generate magnetic fields on a scale large compared with that of the convective cells --- see, for example, the spherical shell simulations of \cite{Olson1999}, \cite{Christensen2006}, \cite{Soderlund2012}, and the plane layer computations of  \cite{Stellmach2004}. However, relating the findings of numerical models to dynamos in the convective cores of rapidly rotating bodies, such as planets, is not entirely straightforward. In computational models, it is not currently feasible to achieve values of the Ekman number ($\Ek$, a measure of  the viscous to the Coriolis force) 
smaller than $\Ek = O(10^{-6})$, whereas in the Earth, for example, $\Ek = O(10^{-15})$. The horizontal extent of convective cells, which depends on the Ekman number as $\Ek^{1/3}$, is therefore expected to be much smaller in nature than in the numerical models; this has important consequences for magnetic field generation \citep[e.g.][]{Jones2000b}. For rapidly rotating planets, the magnetic Reynolds numbers ($\Rm$, the ratio of Ohmic diffusion
time to induction timescale) are expected to be of the order of \mbox{$10^3-10^5$} at the system size; calculated on the small convective scale though, $Rm$ is much less than unity, \ie Ohmic diffusion acts much faster than magnetic induction. In this case, 
a large-scale magnetic field can 
still be generated by the small-scale convective vortices if they act collectively to produce a mean-field $\alpha$-effect \citep{Childress1972, Soward1974}. However, the large-scale magnetic field sustained by this process tends to be spatially uniform \citep[e.g.][]{Favier2013b}, unlike the observed geomagnetic field. In computational models, which necessarily have to consider much higher values of $\Ek$ than the true planetary values, the fundamental problem of very small $Rm$ on the convective scale is therefore implicitly avoided. 
An important challenge of planetary dynamo theory is thus to explain the generation of system-size magnetic fields of strong amplitude and complex spatio-temporal variations, while $Rm$ at the convective scale is smaller than unity.

One plausible solution to this problem is that the generated magnetic field strongly modifies the convective flows
such that the convective scale increases, as predicted by the linear theory of magnetoconvection \citep{Chandrasekhar61}. 
The influence of the magnetic field on the convective flow, and in particular on its lengthscale, has indeed been observed in
a number of dynamo simulations in which strong magnetic fields are sustained \citep[\eg][]{Stellmach2004, Tak08, Hor10, HC_2016}.
In this paper we explore an alternative solution based on a hydrodynamical argument:
in rapidly rotating non-magnetic convection, the small-scale convective vortices may transfer part of their energy to larger-scale flows; if $Rm$ is sufficiently high, based on this increased scale, then the dynamo could operate at these larger scales. The possible formation of large-scale flows is therefore of great interest for the dynamics of planetary interiors. Computationally, this represents a challenging problem, with large domains required that can accommodate many convective cells, together with any large-scale structure that may emerge; as such, studies to date have concentrated on the computationally economical planar geometry. Over the last few years, a number of independent numerical studies of plane layer, non-magnetic, rapidly rotating convection --- both Boussinesq and compressible --- have demonstrated how large-scale vortices (LSVs) can form through the long-term concerted action of the Reynolds stresses resulting from the small-scale convective cells \citep{Chan07, Kapyla11, Rubio14, Favier2014, Guervilly2014}. The LSVs are long-lived, box-size, depth-invariant vortices, which form by the clustering of small-scale convective vortices; their horizontal flows are of much larger amplitude than the underlying convective flows. 

\begin{figure}
\centering
  \includegraphics[clip=true,width=10cm]{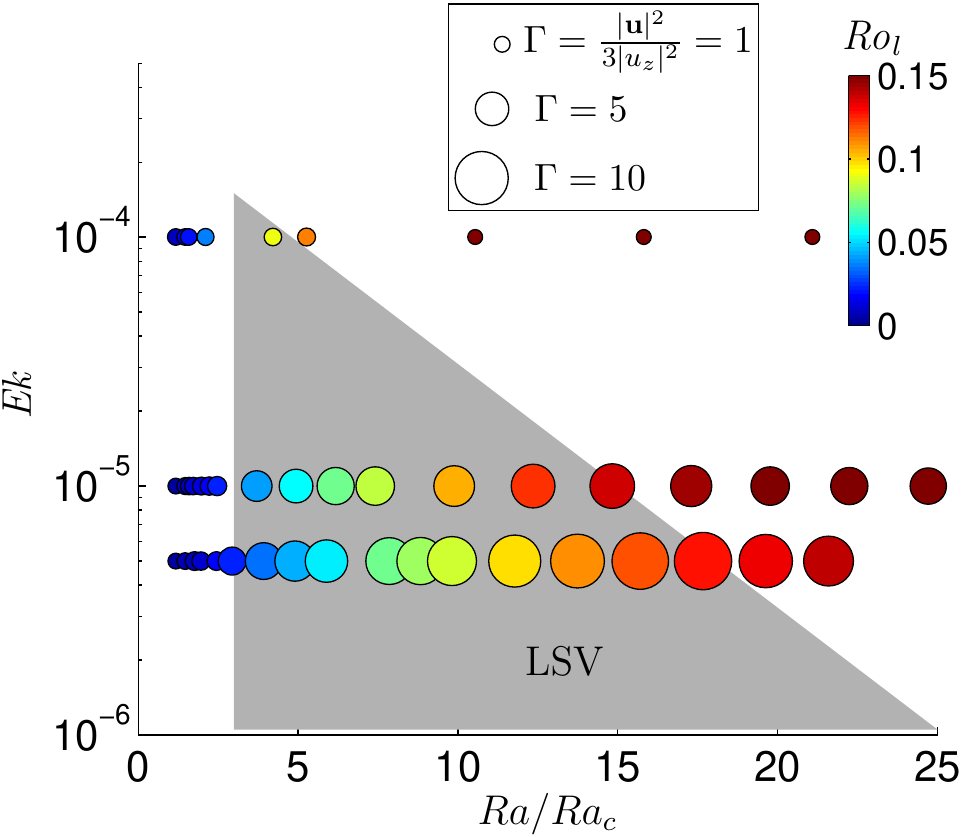}
  \caption{Location of the hydrodynamical simulations of \citet{Guervilly2014} in the parameter space
  $(\Ek,\Ra/\Ra_c)$, for $\Pran=1$ and an aspect ratio of $1$. The colour scale gives the value of the
  local Rossby number, $\Ro_l$, and the area of the circle is proportional to $\Gamma$, which is a measure
  of the relative strength of the horizontal flows. The grey area indicates the region where LSVs form.}
\label{fig:diagram_LSV}
\end{figure}

Figure~\ref{fig:diagram_LSV}, based on the results of \citet{Guervilly2014}, shows the domain of existence of LSVs  for rotating, plane layer, Boussinesq convection, in the parameter space ($\Ek, \Ra/\Ra_c$); the Rayleigh number, $\Ra$, measures the ratio of buoyancy driving to dissipative effects, with $\Ra_c$ the critical value at the onset of convection. 
The area of the circles provides a measure of the relative amplitude of the LSVs,  quantified by the ratio $\Gamma=|\vel|^2/3|u_z|^2$, where $|\vel|$ is the root mean square (rms) value of the total flow and $|u_z|$ is the rms value of the vertical flow. Since the flows in LSVs are essentially horizontal, they are characterised by values of $\Gamma$ larger than unity. 
The colour of the circles denotes the value of the local Rossby number, defined as $\Ro_l = |u_z|/(2\Omega l)$, where $\Omega$ is the rotation rate
and $l$ is the typical horizontal lengthscale of the convection \citep[\eg][]{Christensen2006}. 
$\Ro_l$ is an inverse measure of the rotational constraint on the convective flow; it thus increases with $\Ra$.
There are two essential conditions for the formation of an LSV, highlighted by figure~\ref{fig:diagram_LSV}. One is that the convective flows must be sufficiently energetic to cluster; for the parameter values considered in \citet{Guervilly2014}, this may be expressed by the condition $\Ra /\Ra_c \gtrsim 3$. The other is that the convective flows are rotationally constrained and anisotropic, \ie narrow in the horizontal directions and tall in the vertical direction; this may be expressed by the condition $\Ro_l \lesssim 0.1$. LSVs thus appear for low Ekman numbers and large Reynolds numbers ($\Rey$, the ratio of the viscous timescale to the convective turnover time), precisely the conditions under which convection takes place in planetary cores. LSVs could therefore be good candidates to drive planetary dynamos if they can efficiently generate magnetic fields on scales comparable with or larger than that of the LSVs themselves. 

The question of whether LSVs can indeed drive a dynamo was addressed in the short paper by \citet{Guervilly2015}.  By extending the hydrodynamic study of \citet{Guervilly2014}, it was found that rotating convection in the presence of LSVs can indeed generate a magnetic field with a significant large-scale component. The field is concentrated in the shear layers surrounding the LSVs and is mainly horizontal. A coherent mean (\ie horizontally averaged) magnetic field is also maintained by the flow. This large-scale dynamo process operates only for a range of magnetic Reynolds numbers; $\Rm$ must be large enough for dynamo action to ensue, but small enough that a small-scale magnetic field cannot be permanently sustained by the convective flows. The latter is an essential condition for this particular type of dynamo, since small-scale magnetic fields appear to suppress systematically the formation of the LSV. Indeed, the ability of a small-scale field to disrupt large-scale coherent flows would seem to be a fairly robust characteristic of magnetohydrodynamic turbulence; for example, in a two-dimensional $\beta$-plane model, \citet{Tobias07} found that small-scale fields --- resulting from the distortion of a very weak large-scale field --- suppress the generation of the zonal flows that would otherwise form spontaneously. In the large-scale dynamos considered by \citet{Guervilly2015}, the influence of the small-scale magnetic field varies in time, resulting in sizeable temporal oscillations of the kinetic and magnetic energies. A large-scale magnetic field is first generated by the joint action of the LSV and the smaller-scale convective flow; this large-scale field is then distorted by the convective flows into a small-scale field, which, subsequently, quenches the LSV and triggers the decay of the entire field; once the field is small enough, however, the LSV is able to regenerate and the cycle starts again. Since LSVs consist essentially of horizontal flows, they cannot of themselves act as dynamos \citep{Zeldovich1957}; nonetheless, for convenience, we shall refer to this type of dynamo as an `LSV dynamo', even though it does not rely solely on the LSV.

The present paper builds extensively on \citet{Guervilly2015} by exploring the LSV dynamo in depth. We concentrate on the lowest value of $\Ek$ considered in \citet{Guervilly2014}, thus allowing a wide-ranging exploration in $\Ra$ and $\Rm$. The goals of our study are: (i) to investigate in detail how the LSV dynamo mechanism operates; (ii) to determine the parameter region in which this dynamo operates, in order to determine its relevance for planetary dynamos; and (iii) to explain the mechanism by which the small-scale magnetic field suppresses the LSV.

The layout of the paper is as follows. The mathematical formulation of the problem is given in \S\,\ref{sec:Math}. In \S\,\ref{sec:TTD}, we consider the three very different types of dynamo that can exist in rapidly rotating, plane layer convection, and describe where in parameter space each may be found. Through the application of spectral filters to the convective flows, the key ingredients of the LSV dynamo mechanism are presented in \S\,\ref{sec:Mechanism}. The means by which the LSV can be suppressed and the resulting temporal evolution of the dynamo are described in \S\,\ref{sec:Suppression}. A concluding discussion is contained in \S\,\ref{sec:ccl}.

\section{Mathematical formulation}
\label{sec:Math}

\subsection{Governing equations and boundary conditions}\label{sec:equations}
We consider a three-dimensional Cartesian model of rotating convection for an electrically conducting Boussinesq fluid. Motions are driven by an initially uniform temperature gradient, imposed by fixing the temperature difference \mbox{$\Delta T$} between the top and bottom boundaries. The box depth is $d$. The horizontal dimensions of the computational domain are equal in the $x$ and $y$ directions, with the ratio of horizontal to vertical dimensions denoted by $\lambda$. The acceleration due to gravity is constant, \mbox{$\vect{g} = - g \vect{e}_z$}. The rotation vector is \mbox{$\Omega \vect{e}_z$}. The fluid has kinematic viscosity $\nu$, thermal diffusivity $\kappa$, magnetic diffusivity $\eta$, density $\rho$, thermal expansion coefficient $\alpha$, and magnetic permeability $\mu_0$, all of which are constant. We solve the momentum, temperature 
and magnetic induction equations in dimensionless form, obtained by scaling lengths with $d$, times with \mbox{$1/(2\Omega)$}, temperature with \mbox{$\Delta T$}, and magnetic field with \mbox{$2\Omega d(\rho \mu_0)^{1/2} $}. The resulting system of dimensionless governing equations is
\begin{equation}
	\pdt{\vel} + \vel \bcdot \bnabla \vel + \vect{e}_z \times \vel =
	- \bnabla p
	+ \frac{\Ra \Ek^2}{\Pran} \theta \vect{e}_z + \Ek \nabla^2 \vel 
        + \pleft \bnabla \times \B \pright \times \B, 
	\label{eq:u}
\end{equation}
\begin{equation}	
         \bnabla \bcdot \vel = 0 ,
\end{equation}
\begin{equation}
	 \pdt{\theta} + \vel \bcdot  \bnabla \theta - u_z = \frac{\Ek}{\Pran} \nabla^2 \theta ,
	\label{eq:theta}
\end{equation}
\begin{equation}
	\pdt{\B} = \bnabla \times \pleft \vel \times \B \pright + \frac{\Ek}{\Pm} \nabla^2 \B ,
	\label{eq:B}
\end{equation}
\begin{equation}
	\bnabla \bcdot \B = 0,
	\label{eq:divB}
\end{equation}
where \mbox{$\vel=(u_x,u_y,u_z)$} is the velocity, $p$ the pressure, $\theta$ the temperature perturbation relative to a linear background profile, and \mbox{$\B=(B_x,B_y,B_z)$} the magnetic field. The dimensionless parameters are the Rayleigh number,
\begin{equation}
	\Ra = \frac{\alpha g \Delta T d^3}{\kappa \nu},
\end{equation}
the Ekman number,
\begin{equation}
	\Ek = \frac{\nu}{2\Omega d^2},
\end{equation}
and the thermal and magnetic Prandtl numbers,
\begin{equation}
	\Pran = \frac{\nu}{\kappa}, \quad \Pm = \frac{\nu}{\eta}.
\end{equation}

We assume that all variables are periodic in the horizontal directions. In the vertical direction, 
the upper and lower boundaries are taken to be perfect thermal and electrical conductors, impermeable and stress-free, \ie
\begin{align}
 	\theta &= 0 \quad  \mbox{at } z=0,1;
	\label{eq:BCtheta}
	\\
	\frac{\partial B_x}{\partial z} &=  \frac{\partial B_y}{\partial z} = B_z =0 \quad  \mbox{at } z=0,1 ;
	\label{eq:BC_B}
	\\
	\frac{\partial u_x}{\partial z} &=  \frac{\partial u_y}{\partial z} = u_z =0 \quad  \mbox{at } z=0,1 .
	\label{eq:BCu}		
\end{align}

It is worth noting that the implementation of stress-free boundary conditions provides the best opportunity for the development of horizontal flows of large amplitude. For the Ekman numbers considered here, no-slip boundary conditions inhibit the formation of LSVs \citep{Stellmach2014}.

On occasion, we shall refer to results obtained for the \textit{kinematic} dynamo problem. This is governed by equations~\eqref{eq:u}--\eqref{eq:divB}, but with the Lorentz force (the final term on the right hand side of \eqref{eq:u}) omitted; there is then no feedback from the magnetic field onto the flow, and the problem becomes linear in the magnetic field $\B$.

\subsection{Numerical method}
Equations~(\ref{eq:u})--(\ref{eq:divB}) are solved numerically using a parallel pseudospectral code developed by \citet{Cattaneo03}. The temperature perturbation and each component of the velocity are transformed from configuration space (containing \mbox{$N_x \times N_y \times N_z$} collocation points) to phase space
by a discrete Fourier transform of the form
\begin{equation}
 f(x,y,z) =  \sum\limits_{k_x}  \sum\limits_{k_y}  \sum\limits_{k_z} \hat{f} (k_x,k_y,k_z)
	      \exp (2\pi i k_x x) \exp (2\pi i k_y y) \phi (\pi k_z z) + \mathrm{c.c.}, 
\end{equation}
where $f$ and $\hat{f}$ are the functions in configuration and phase spaces respectively and $\mathrm{c.c.}$ denotes complex conjugate.
The function $\phi(s)$ is governed by the boundary conditions: $\phi(s)=\cos(s)$ for $u_x$, $u_y$, $B_x$ and $B_y$, whereas $\phi(s) = \sin(s)$ for $u_z$, $B_z$ and $\theta$.
Further details concerning the numerical methods can be found in \citet{Cattaneo03}. 
Table~\ref{tab:res} gives the parameter values and numerical resolution of the simulations discussed in this paper.

\begin{table}
 \begin{center}
  \begin{tabular}{lcccc}
    $\Ra$  & \hspace{0.3cm} & $\Pm$ &   \hspace{0.3cm} &$N_x \times N_y \times N_z$ \\[3pt]
    \hline
    $1.2\times10^8$ & & $0.2$, $0.5$, $1$, $2.5$ & &$256\times256\times128$ \\
    $1.5\times10^8$ & & $2.5$ & &$256\times256\times128$ \\
    $1.8\times10^8$ & &$0.2$, $0.5$, $1$, $2.5$ & & $256\times256\times128$ \\
    $2\times10^8$ & & $2.5$ & & $256\times256\times128$ \\
    $2.5\times10^8$ & & $0.2$, $0.5$, $1$, $2$, $2.5$ & & $256\times256\times128$ \\
    $3\times10^8$ & & $0.2$, $0.5$, $1$, $2$, $2.5$ & & $256\times256\times128$ \\
    $4\times10^8$ & & $0.1$, $0.2$, $0.5$, $1$, $2.5$ & & $256\times256\times256$ \\
    $5\times10^8$ & & $0.1$, $0.2$, $0.3$, $0.5$, $1$, $2.5$ & & $256\times256\times256$ \\
    $1\times10^9$ & &$0.1$, $0.2$, $0.5$ & & $512\times512\times256$ \\
  \end{tabular}
  \caption{Summary of the parameter values and numerical resolution for the simulations performed at $\Ek=5\times10^{-6}$ and $\aspect=1$.}
  \label{tab:res}
 \end{center}
\end{table}

\subsection{Output parameters for the flow and field}
The convective dynamo problem is formulated unambiguously by the input parameters described in \S\,\ref{sec:equations}, together with the initial conditions. It is helpful also to introduce additional output parameters that characterise the resulting velocity and magnetic field.

We define the Reynolds number in terms of the vertical velocity, which is a representative velocity of the convective flow, and the box depth, \ie
\begin{equation}
	\Rey = \frac{d \sqrt{\langle u_z ^2\rangle_{xyz}}}{\nu},
\end{equation}
where $\langle \, \cdot \, \rangle_a$ denotes an average over the direction $a$. The magnetic Reynolds number is defined by $\Rm=\Rey \Pm$.

We define the energy spectrum of the horizontal velocity, $\vect{u}_h = (u_x,u_y,0)$, by
\begin{equation}
  E_u^h (k_h) = \frac{1}{2} \sum\limits_{k_z}  C | \hat{\vect{u}}_h (k_x,k_y,k_z) |^2 ,
	\label{eq:Eh}
\end{equation}
where $k_h=\sqrt{k_x^2+k_y^2}$ is the horizontal wavenumber
and $C=(1+\delta_{k_x 0}\delta_{k_y 0})(1+\delta_{k_z 0})$.
Similarly, the energy spectrum of the vertical velocity, $(0,0,u_z)$, is defined by
\begin{equation}
	E_u^v (k_h) = \frac{1}{2} \sum\limits_{k_z} C | \hat{u}_z (k_x,k_y,k_z)|^2.
	\label{eq:Ev}
\end{equation} 
The energy spectra are obtained by binning the energy into rings of radius $k_h$ with $\Delta k_h = 1/\lambda$. In the same way, we denote the energy spectra of the horizontal magnetic field, $\B_h = (B_x,B_y,0)$, and the vertical magnetic field, $(0,0,B_z)$, by $E_b^h$ and $E_b^v$ respectively. 

Finally, the integral horizontal wavenumber of the convective flow is defined as
\begin{equation}
   {k_{u}^{v}} = 
               \frac{ \sum\limits_{k_x,\, k_y,\, k_z}   \sqrt{k_x^2+k_y^2}  C | \hat{u}_z(k_x, k_y, k_z) |^2}
		   {\sum\limits_{k_x,\, k_y,\, k_z}  C |\hat{u}_z(k_x, k_y, k_z) |^2} ,
\label{eq:kh}
\end{equation}
where $k_u^v$ is also averaged in time. We use a similar formula to calculate the integral wavenumbers of  $\B_h$ (denoted by $k_b^h$) and $B_z$ ($k_b^v$).

\section{Three types of dynamo}
\label{sec:TTD}

\begin{figure}
\centering
  \includegraphics[clip=true,width=13cm]{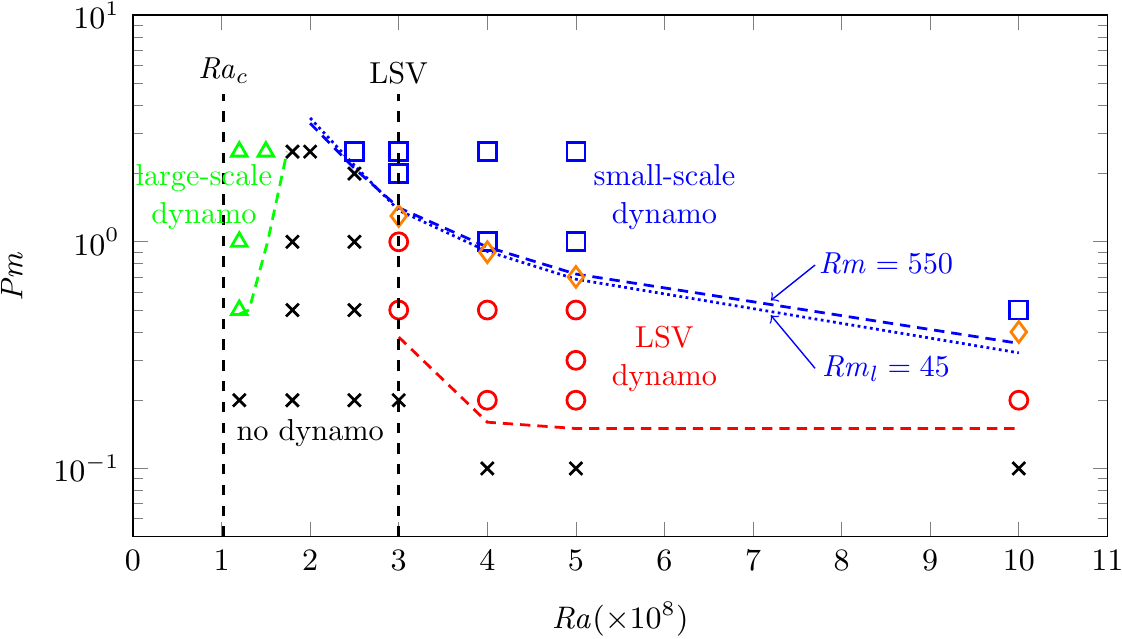}
  \caption{
Location in $(\Ra,\Pm)$ space of the simulations that either successfully generate dynamos (coloured triangles, squares and circles), or that fail to produce a dynamo (black crosses): $\Ek=5\times10^{-6}$ and $\aspect = 1$. The transition from the LSV dynamo to the small-scale dynamo, determined by the location at which $k_b^h=k_u^v$, is represented by diamonds. $\Rm$ is the magnetic Reynolds number defined with the vertical velocity and the box depth; $\Rm_l$ is defined with the vertical velocity and the convective scale. The red and green dashed lines represent, respectively, the thresholds for the LSV dynamo and for the large-scale dynamo near the onset of convection.}
\label{fig:diagram}
\end{figure}

\begin{figure}
\centering
  \subfigure[]{\label{fig:wz_Onsdyn}
  \includegraphics[clip=true,width=0.32\textwidth]{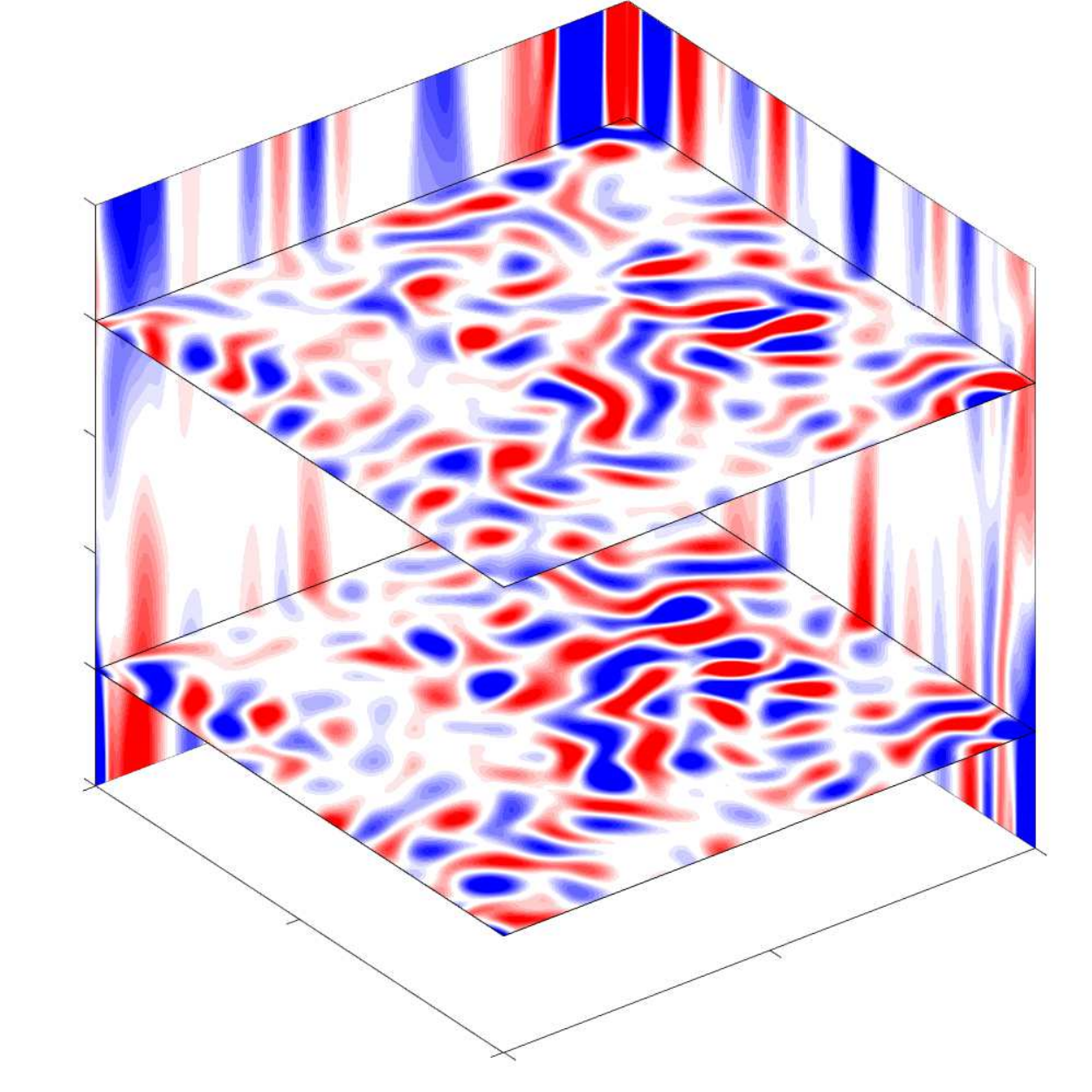}}
  \subfigure[]{\label{fig:wz_LSVdyn}
  \includegraphics[clip=true,width=0.32\textwidth]{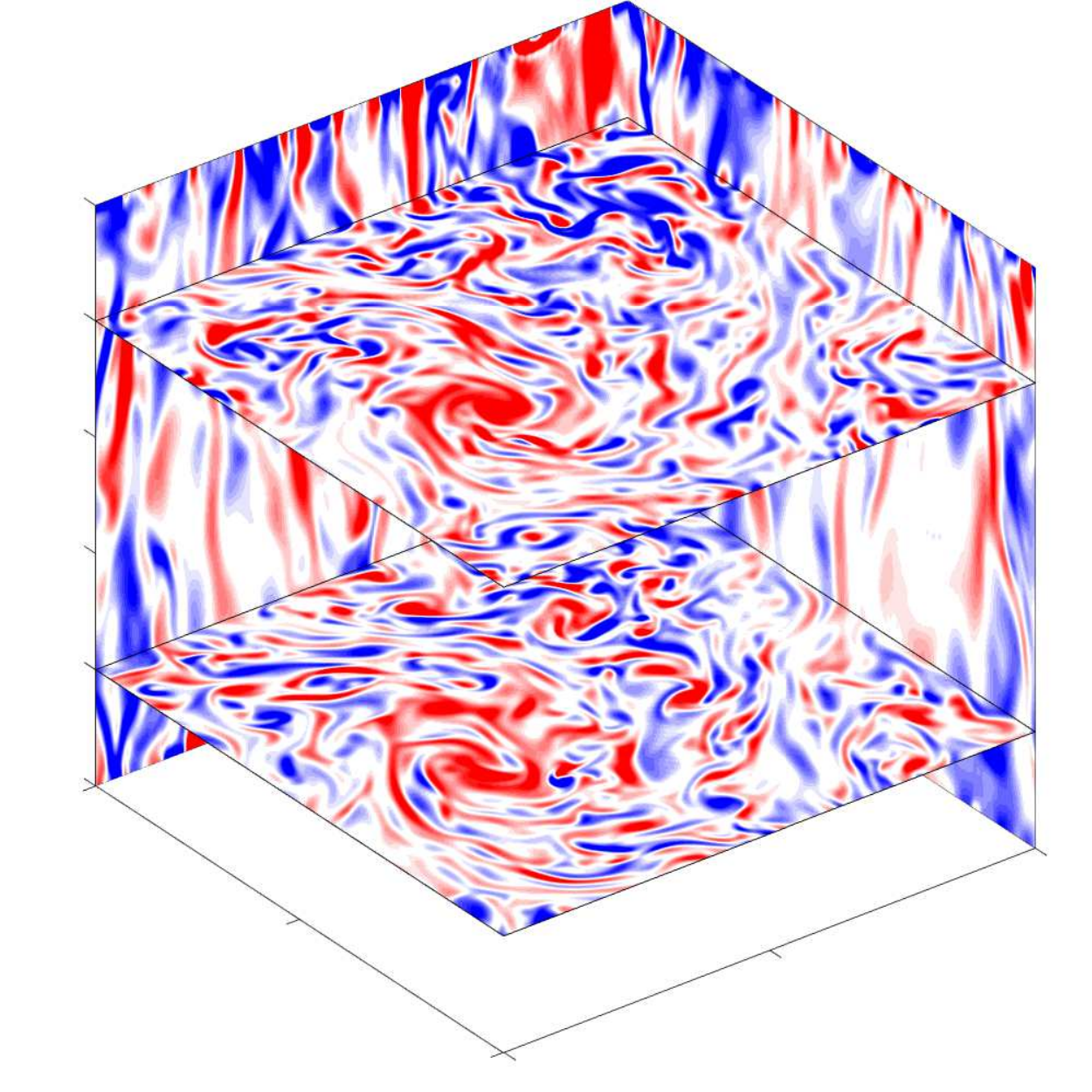}}
  \subfigure[]{\label{fig:wz_smalldyn}
  \includegraphics[clip=true,width=0.32\textwidth]{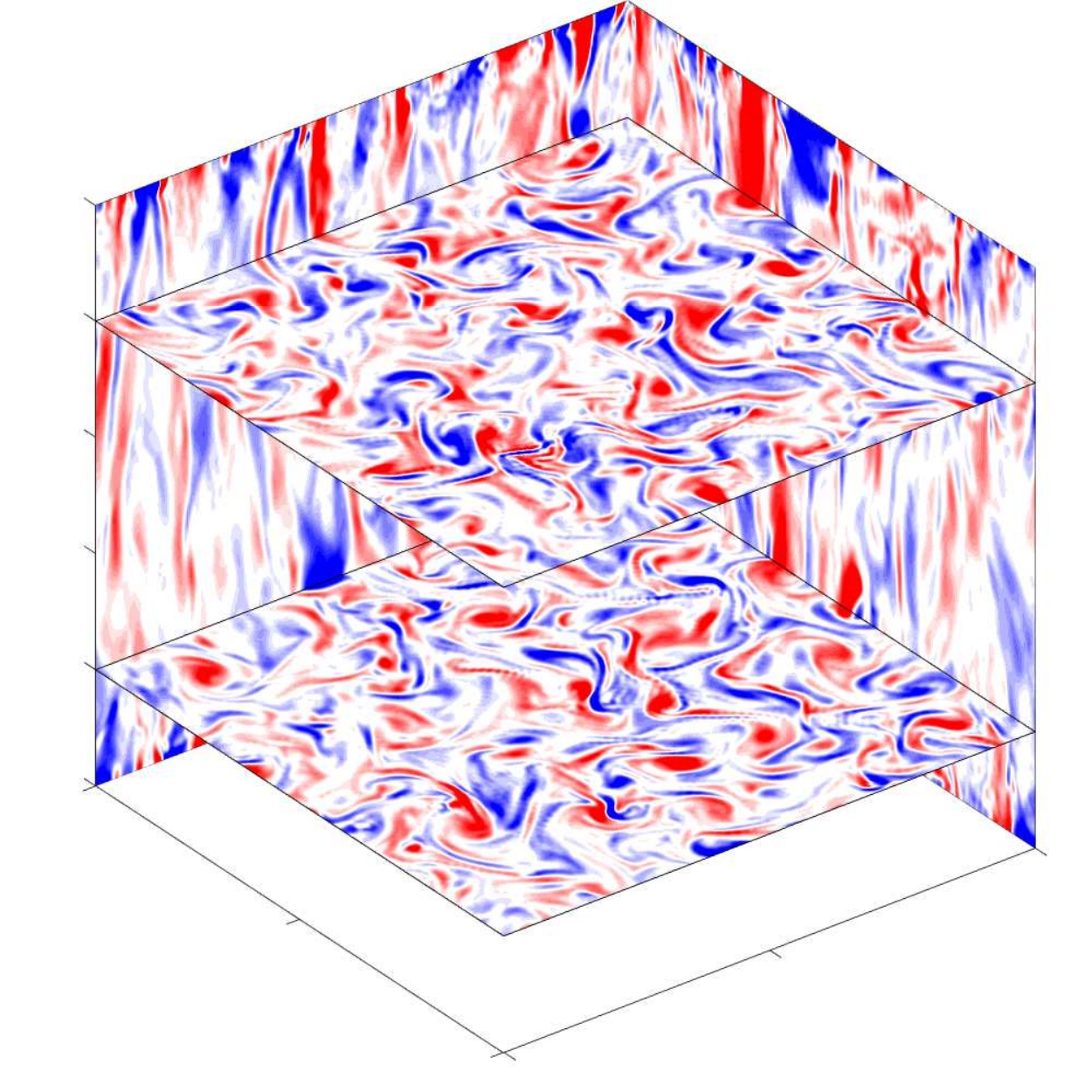}}
    \subfigure[]{\label{fig:Bx_Onsdyn}
  \includegraphics[clip=true,width=0.32\textwidth]{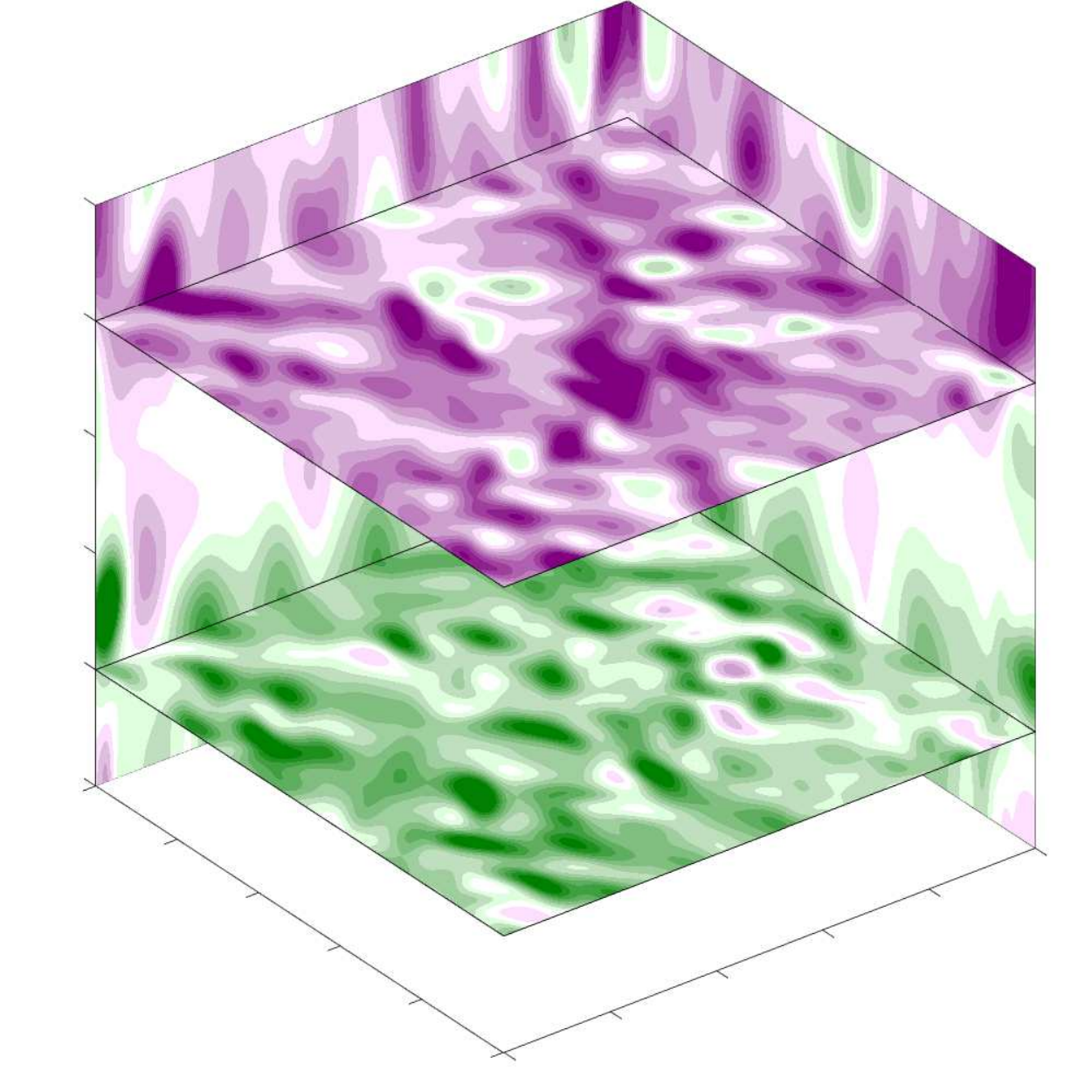}}
    \subfigure[]{\label{fig:Bx_LSVdyn}
  \includegraphics[clip=true,width=0.32\textwidth]{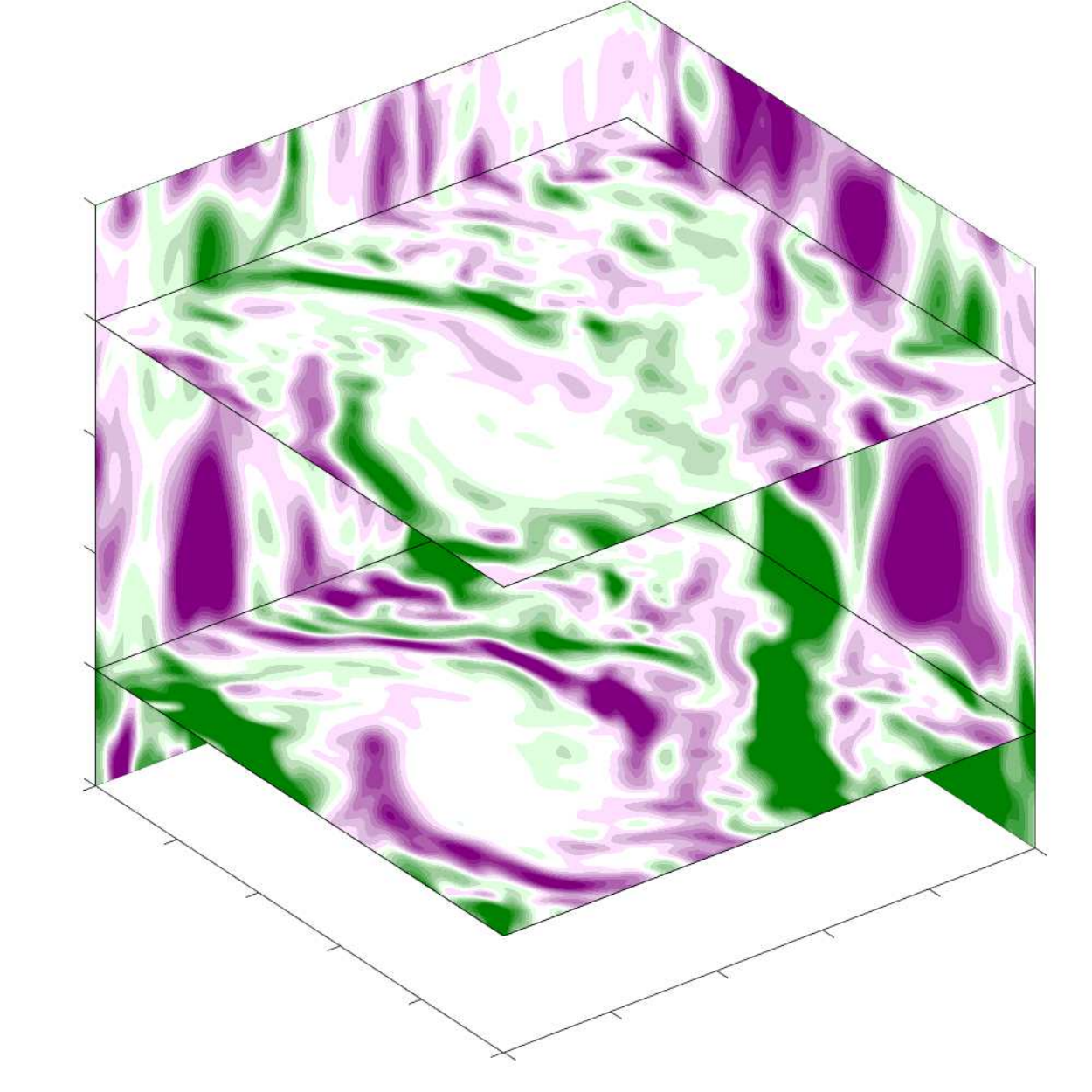}}
  \subfigure[]{\label{fig:Bx_smalldyn}
  \includegraphics[clip=true,width=0.32\textwidth]{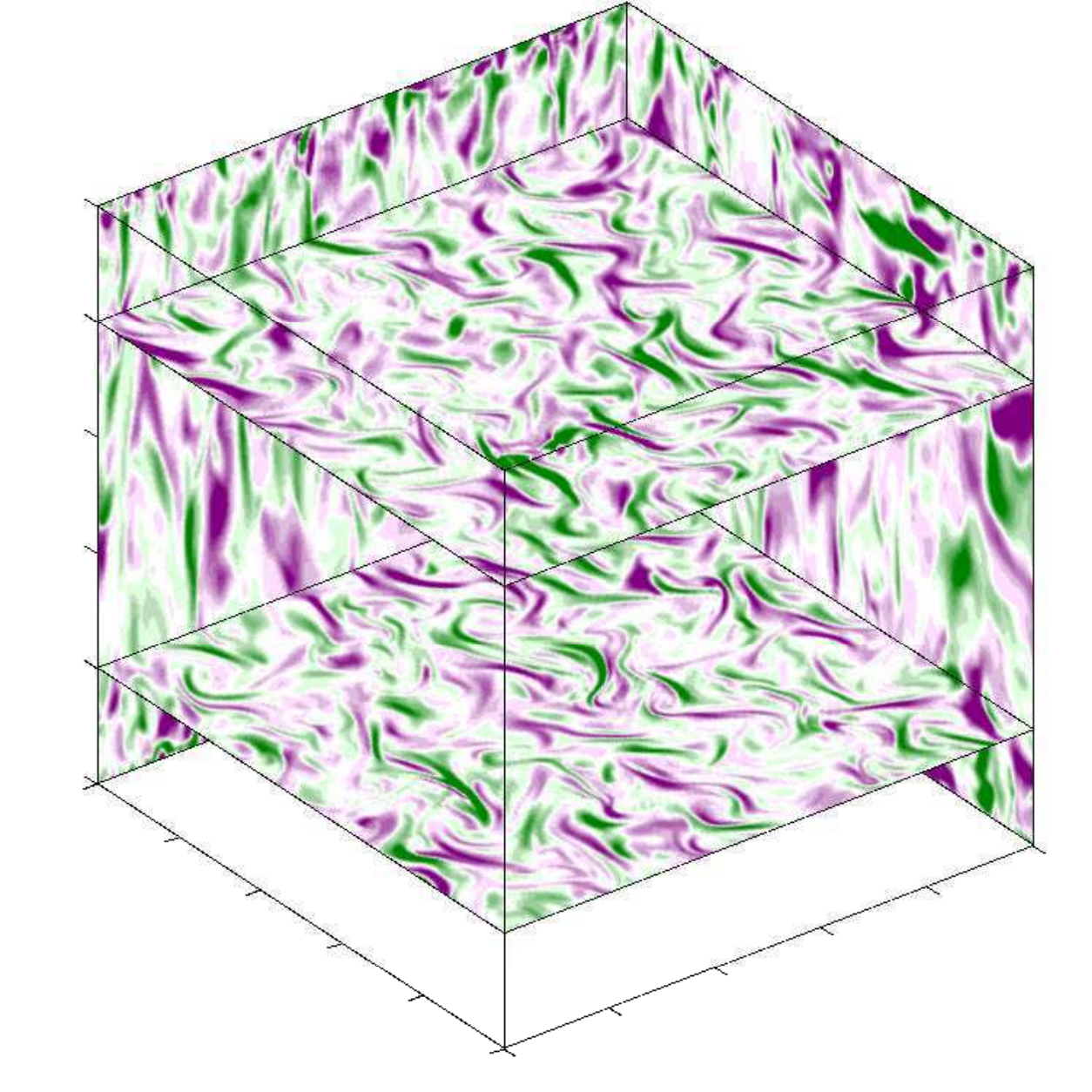}}
  \caption{Snapshots of the horizontal and vertical cross-sections of (a)-(c) the axial vorticity
  and (d)-(f) $B_x$, for three of the dynamo simulations marked in figure~\ref{fig:diagram}:
  (a) and (d) $\Ra=1.2\times10^8$ and $\Pm=0.5$ (large-scale dynamo close to the onset of convection);
  (b) and (e) $\Ra=5\times10^8$ and $\Pm=0.2$ (LSV dynamo);
  (c) and (f) $\Ra=5\times10^8$ and $\Pm=2.5$ (small-scale dynamo).}
\label{fig:wz_Bx}
\end{figure}

In this paper we consider rapidly rotating convection at fixed values of the Ekman number, Prandtl number and aspect ratio ($\Ek=5\times10^{-6}$, $\Pran=1$ and $\aspect=1$), for a range of Rayleigh numbers and magnetic Prandtl numbers. The location in $(\Ra, \Pm)$ parameter space of the simulations discussed is shown in figure~\ref{fig:diagram}. Also marked on the figure are the critical Rayleigh number for the onset of convection in the asymptotic limit of small $\Ek$, $\Ra_c \approx 1.02\times 10^8$ \citep{Chandrasekhar61}, and the threshold Rayleigh number above which LSVs form in non-magnetic convection, given by $\Ra \approx 3 \times 10^8$. It is also worth noting that for this value of $\Ek$, LSV formation ceases for $\Ra \gtrsim 1.8 \times 10^9$, as shown by \cite{Guervilly2014} (see figure~\ref{fig:diagram_LSV}); although hydrodynamical simulations are feasible at such high $\Ra$, dynamo simulations become impracticable for $\Ra \gtrsim 10^9$.

In the parameter regime covered by figure~\ref{fig:diagram}, we have identified three very distinct types of dynamo. This feature is illustrated clearly in figure~\ref{fig:wz_Bx}, which shows the axial vorticity and the $x$-component of the magnetic field for representative cases of each type of dynamo. In the following subsections we explore in some detail the characteristics of the flow and magnetic field for these three dynamo mechanisms.

\subsection{Large-scale dynamo near the onset of convection}

\begin{figure}
\centering 
  \subfigure[]{\label{fig:LSdynamo_b}
  \raisebox{1cm}{\includegraphics[clip=true,width=0.40\textwidth]{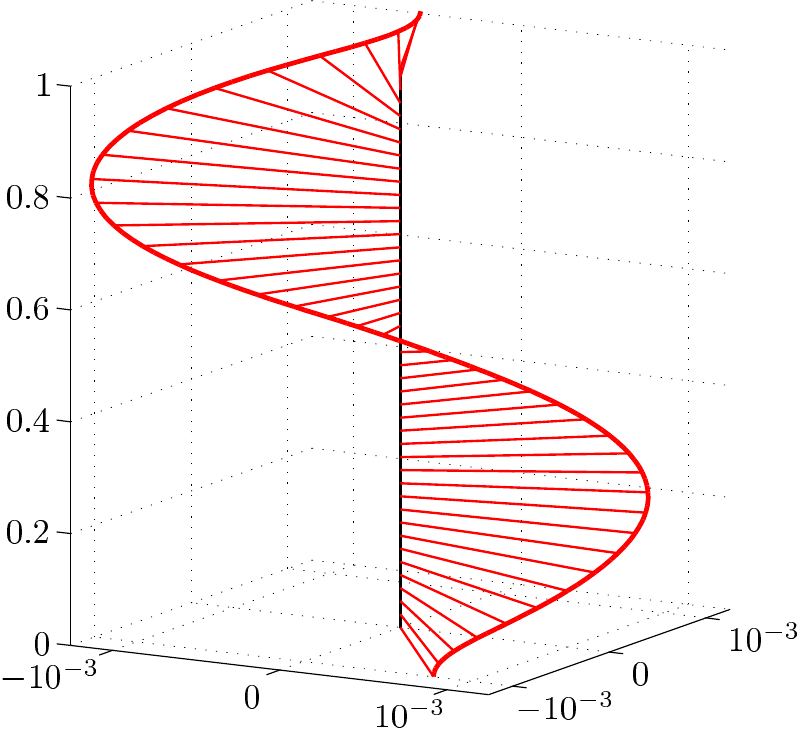}}}
  \subfigure[]{\label{fig:LSdynamo_c}
  \includegraphics[clip=true,width=0.54\textwidth]{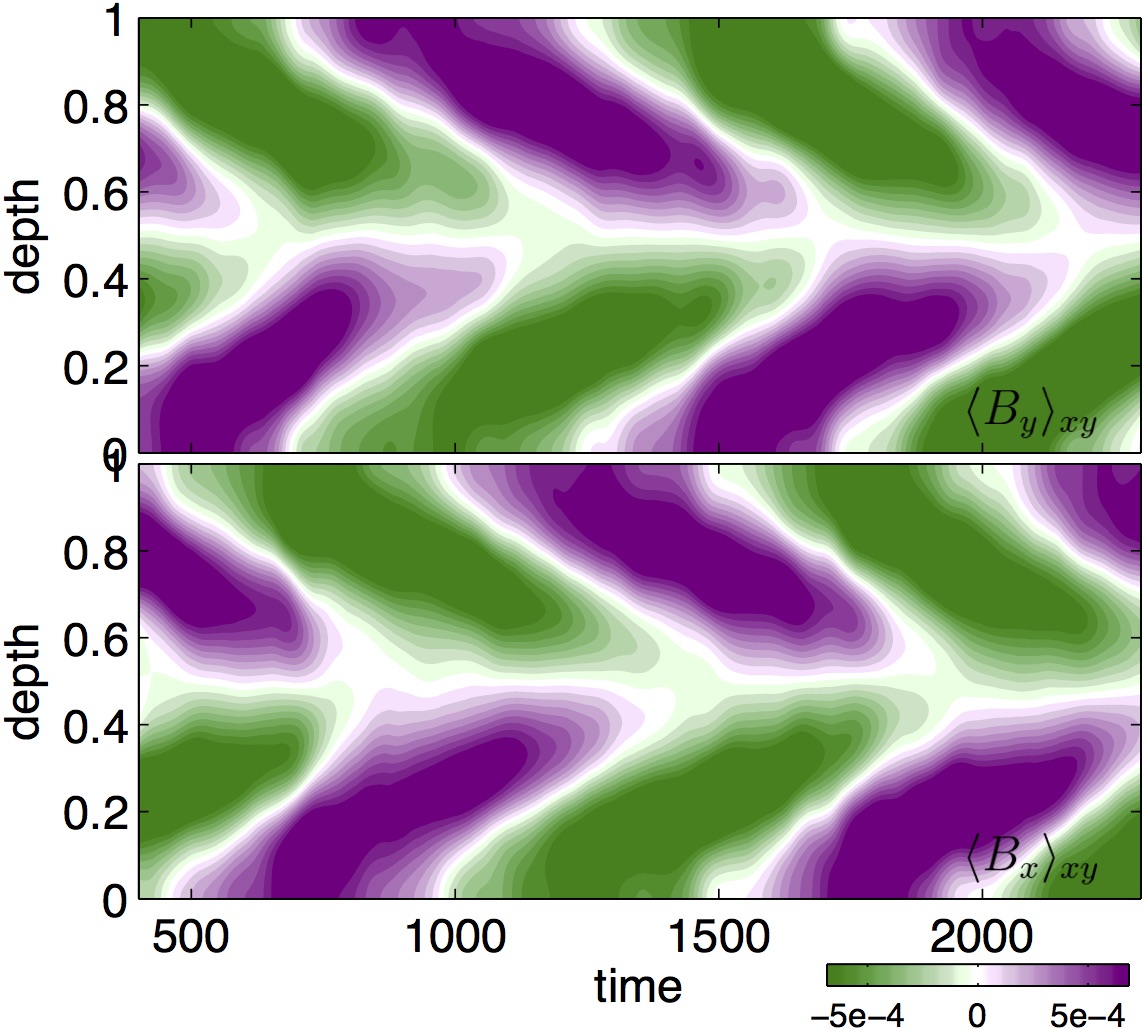}}
  \caption{
  Mean magnetic field produced by the dynamo near the onset of convection ($\Ra=1.2\times10^8$ and $\Pm=0.5$).
  (a) Snapshot of the vertical structure of the mean magnetic field 
  $(\langle B_x \rangle_{xy}, \langle B_y \rangle_{xy},0)$;
  (b) space-time diagram of $\langle B_x \rangle_{xy}$ and $ \langle B_y \rangle_{xy}$.}
\label{fig:LSdynamo}
\end{figure}

For Rayleigh numbers just above $\Ra_c$, the convection takes the form of elongated columns, which have a small horizontal cross-section and are aligned with the rotation axis. In the non-magnetic case, the integral convective wavenumber (defined by (\ref{eq:kh})) is $k_{u}^{v} = 12.6$ for $\Ra=1.2\times10^8$. For this Rayleigh number, the convective flows maintain a dynamo for $\Pm \gtrsim 0.5$, which corresponds to $Rm \gtrsim 24$. The magnetic field is dominated by a mean (\ie horizontally-averaged) component; this is clearly seen in the horizontal cross-section of $B_x$ in figure~\ref{fig:Bx_Onsdyn}. Note that, owing to the magnetic boundary conditions~(\ref{eq:BC_B}), the mean field can have only a horizontal component. This type of large-scale dynamo, which operates near the onset of convection, was first described analytically by \citet{Childress1972} and \citet{Soward1974}, and was later studied through detailed numerical simulations of rotating convection by \citet{Stellmach2004} and by consideration of specific flow planforms by \citet{Favier2013b} and \cite{Calkins2015}. Details of this dynamo are recalled here for comparison with the dynamos found at larger values of $\Ra$. The dynamo process works as a two-scale mechanism, whereby the mean field is generated by the electromotive force (emf) produced by the collective action of small-scale convective columns. Figure~\ref{fig:LSdynamo_b} shows the vertical structure of the mean magnetic field $(\langle B_x \rangle_{xy}, \langle B_y \rangle_{xy}, 0)$, revealing a
spiral staircase structure that is anti-symmetric with respect to the mid-plane. Figure~\ref{fig:LSdynamo_c} shows the space-time diagram of $ \langle B_x \rangle_{xy}$ and $ \langle B_y \rangle_{xy}$. The components of the mean field have a well-defined periodicity corresponding to a clockwise rotation of the entire staircase structure, with the maxima moving in time from the boundaries towards the mid-plane \citep{Stellmach2004, Favier2013b}.

In figure~\ref{fig:diagram}, the green dashed line represents the transition between the regime of this mean-field, large-scale dynamo found close to the onset of convection and a regime in which there is no dynamo action; this boundary was deduced from consideration of the kinematic dynamo growth rates of neighbouring points in parameter space. The most significant, and somewhat counter-intuitive, feature of this dynamo is that increasing the Rayleigh number, and hence the strength of the convection and the magnetic Reynolds number, eventually kills off the dynamo process. Mean-field dynamo action relies crucially on a high degree of spatial and temporal coherence of the small-scale velocity in order to provide an effective emf \citep[see, for example,][]{Courvoisier2009}. As the convection becomes more vigorous, although there is more kinetic energy available to drive a dynamo, the coherence of the convective columns is lost, leading to the failure of the large-scale dynamo \citep{Cattaneo06, HC_2008, Tilgner12}. Critically, this occurs even though the convective vortices remain rotationally constrained. Hence this large-scale dynamo is confined very close to the onset of convection, \ie $\Ra \lesssim 1.5 \Ra_c$ for $\Ek = 5 \times 10^{-6}$. Provided the magnetic Reynolds number is sufficiently high, the convective flows that are unable to generate a mean (large-scale) magnetic field are then able to generate only small-scale fields (\ie fields with a size comparable with or smaller than the convective scale) \citep{Cattaneo06}.

\subsection{Dynamos from LSVs}

For sufficiently large thermal driving, an LSV forms in non-magnetic, rapidly rotating convection; an example is shown in figure~\ref{fig:wz_LSVdyn}. The LSV consists of a concentrated cyclone and a more dilute anticyclone; details of the formation and stability of the LSV are discussed in detail in \cite{Favier2014}, \cite{Guervilly2014}, \cite{Rubio14} and \cite{Stellmach2014}. Starting from such convective flows, two very different types of dynamo can then be generated, depending on the values of $\Rm$, or, equivalently, on $\Pm$ for fixed $\Rey$; in one the LSV is preserved by the magnetic field, whereas in the other it is destroyed.

\begin{figure}
\centering
   \subfigure[]{
  \includegraphics[clip=true,width=0.45\textwidth]{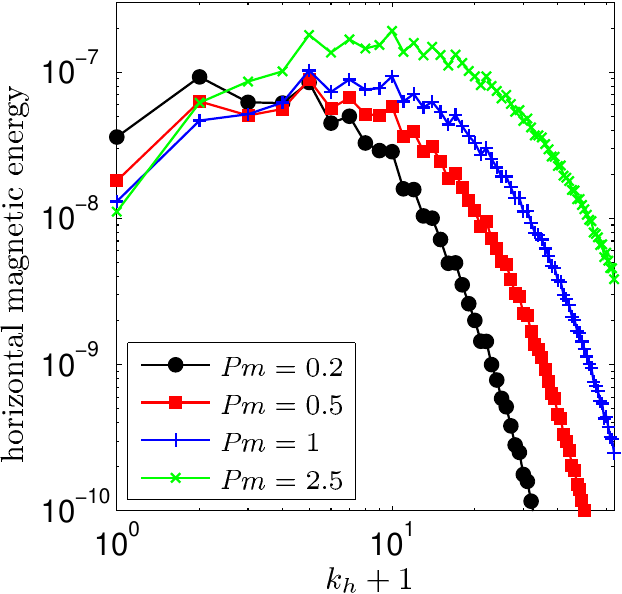}}
  \subfigure[]{
  \includegraphics[clip=true,width=0.45\textwidth]{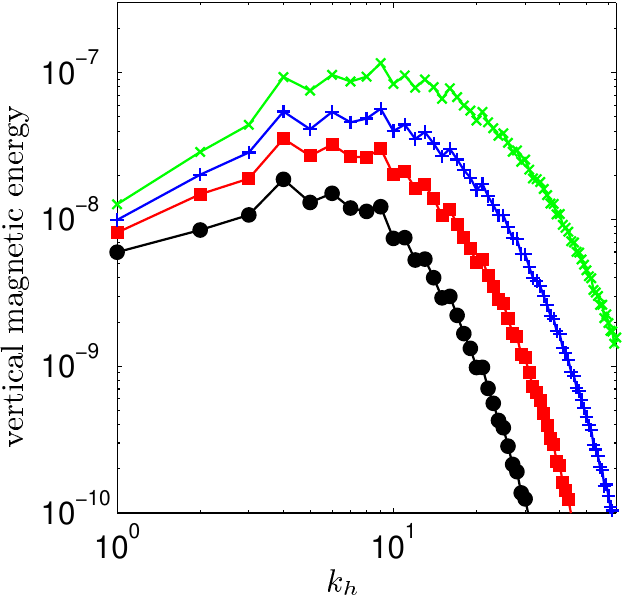}}
  \caption{
  Time-averaged horizontal power spectra of (a) the horizontal magnetic energy $E_b^h$, and
  (b) the vertical magnetic energy $E_b^v$, for different $\Pm$ and $\Ra=5\times10^8$.}
\label{fig:specb_Ra5e8}
\end{figure}

Figure~\ref{fig:specb_Ra5e8} shows the time-averaged horizontal power spectra of the magnetic energies for $\Ra = 5 \times 10^8$ and for different values of $\Pm$. For $\Pm=0.2$, the energy of the horizontal field is dominated by small wavenumbers, $k_h \lesssim 10$; the vertical magnetic field is significantly weaker than the horizontal field for $k_h < 12$. As $\Pm$ is increased beyond unity, the magnetic energy moves towards larger wavenumbers, peaking around the integral convective wavenumber $k_u^v \approx 12.1$. For $\Pm=0.2$, the mean magnetic field (corresponding to the wavenumber $k_h=0$) contains on average $5$\% of the total magnetic energy, while for $\Pm=2.5$, it represents only $0.3$\%.

\begin{figure}
\centering
  \includegraphics[clip=true,width=0.6\textwidth]{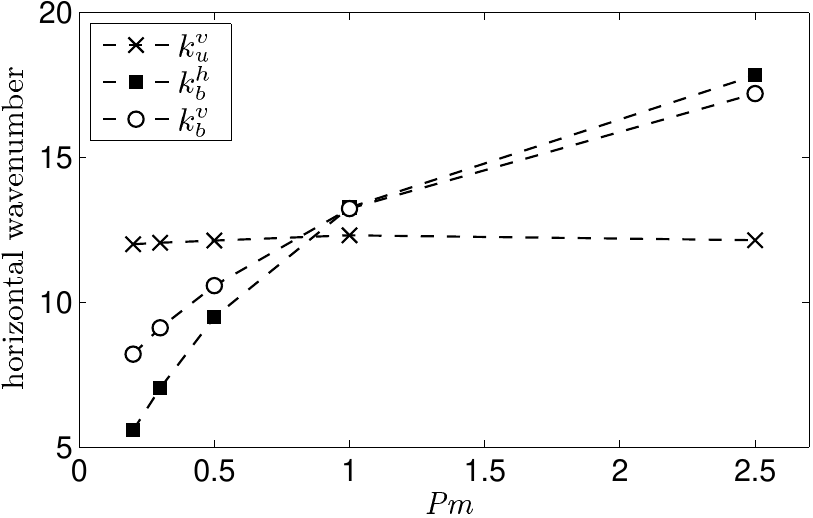}
  \caption{Integral horizontal wavenumbers as a function of $\Pm$ with $\Ra = 5 \times 10^8$.}
\label{fig:scales_Ra5e8}
\end{figure}

Figure~\ref{fig:scales_Ra5e8} shows the integral wavenumbers of the horizontal magnetic field, $k_b^h$, and the vertical magnetic field, $k_b^v$, as a function of $\Pm$ with $\Ra = 5 \times10^8$. For comparison, the integral convective wavenumber, $k_u^v$, is also plotted. As expected from inspection of the magnetic energy spectrum, both $k_b^h$ and $k_b^v$ increase with $\Pm$. For $\Pm \lesssim 1$, the integral magnetic wavenumbers
are smaller than the convective wavenumber, so most of the magnetic energy  
is contained in the large scales. Also in these cases, $k_b^h$ is systematically smaller than $k_b^v$, so the horizontal magnetic field is
 dominated by scales larger than those of the vertical field.
For $\Pm \gtrsim 1$, the integral magnetic wavenumbers are larger than the convective wavenumber, so 
most of the magnetic energy is contained in scales smaller than the convective scale; we refer to these
dynamos as \textit{small-scale} dynamos. In these cases, the integral wavenumbers of the horizontal and vertical magnetic fields are comparable. Hence there is a transition from a dynamo that generates a large-scale magnetic field (i.e.\ an LSV dynamo) to a small-scale dynamo; for this particular Rayleigh number ($\Ra=5\times10^8$), the transition occurs for $\Pm$ in the range $0.5 < \Pm < 1$.
Figure~\ref{fig:diagram} shows the magnetic Prandtl number for this transition (\ie when $k_b^h=k_u^v$) for different values of $\Ra$. Note that the transition between the two dynamo regions is continuous.

The difference in the spatial structure of the magnetic field generated by the LSV dynamo at small $\Pm$ and that of the small-scale dynamo at large $\Pm$ is clearly demonstrated in the horizontal slices of $B_x$ shown in figures~\ref{fig:Bx_LSVdyn} and \ref{fig:Bx_smalldyn}. The magnetic field for $\Pm=0.2$ is organised into large-scale structures, concentrated in horizontal bands localised in the shear layers surrounding the LSV. In each band of concentrated field, the horizontal magnetic field has one main direction. In the core of the cyclone, the field intensity is weak. By contrast, for $\Pm=2.5$, the magnetic field is dominated by small scales; no large-scale organisation of the field is apparent.

\begin{figure}
\centering
  \hspace{0.4cm}
  \subfigure[]{\label{fig:LSVD_Bmean_a}
   \raisebox{1cm}{\includegraphics[clip=true,width=0.40\textwidth]{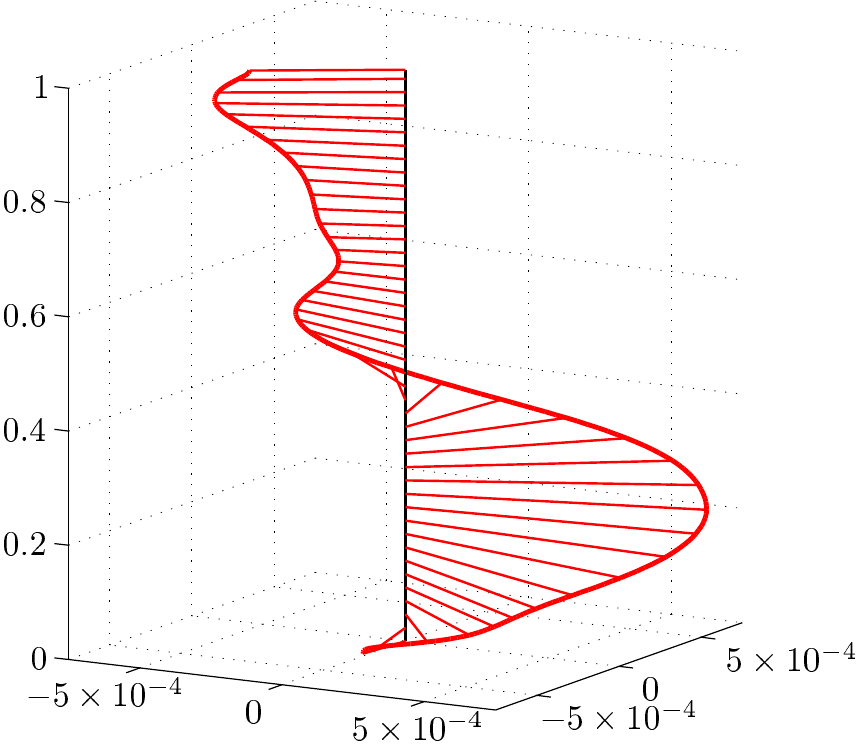}}}
  \subfigure[]{\label{fig:LSVD_Bmean_b}
  \includegraphics[clip=true,width=0.54\textwidth]{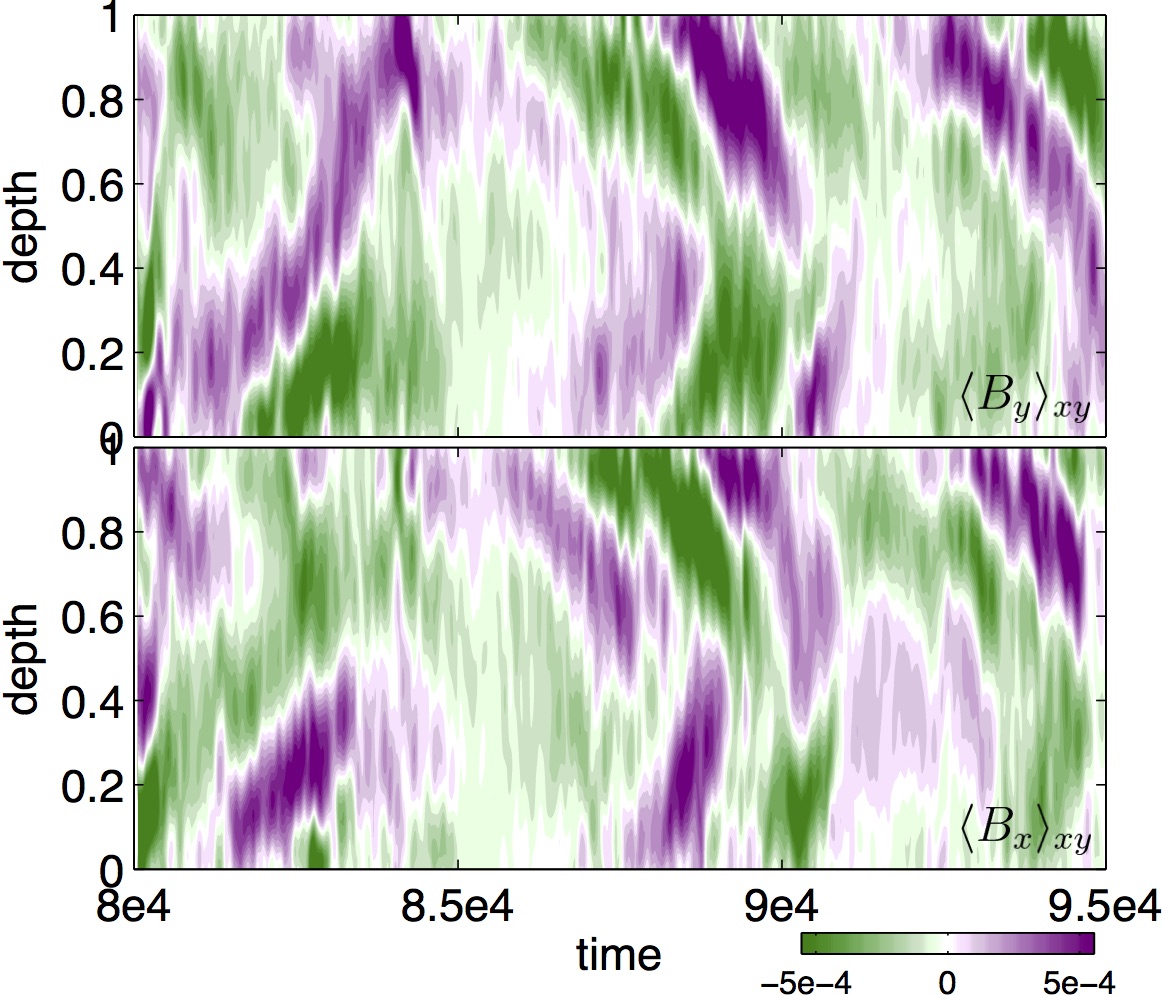}}
  \caption{Mean magnetic field produced by the LSV dynamo ($\Ra=5\times10^8$ and $\Pm=0.2$).
  (a) Snapshot of the vertical structure of $( \langle B_x \rangle_{xy}, \langle B_y \rangle_{xy},0)$;
  (b) space-time diagram of $ \langle B_x \rangle_{xy}$ and $ \langle B_y \rangle_{xy}$.}
  \label{fig:LSVD_Bmean}
\end{figure}

Figure~\ref{fig:LSVD_Bmean_a} shows the vertical structure of the mean magnetic field $( \langle B_x \rangle_{xy},  \langle B_y \rangle_{xy},0)$ for the LSV dynamo at $\Pm=0.2$. It has a large-scale variation along $z$ with a structure similar to a spiralling staircase, but which is more complex than that of the mean field for the large-scale dynamo near the onset of convection (cf.\ figure~\ref{fig:LSdynamo_b}). Figure~\ref{fig:LSVD_Bmean_b} shows the space-time diagram of $ \langle B_x \rangle_{xy}$ and $ \langle B_y \rangle_{xy}$. The direction and amplitude of the mean magnetic field evolves in time in an approximately periodic manner, with a tendency to drift from the boundaries towards the mid-plane. The averaged fields $\langle B_x \rangle_{xy}$ and $ \langle B_y \rangle_{xy}$ are preferentially antisymmetric with respect to the mid-plane. The entire staircase structure tends to rotate clockwise, in a similar manner to that of the large-scale dynamo operating close to the onset of convection (figure~\ref{fig:LSdynamo_c}), but here with a slower rotation rate. 
 
\begin{figure}
\centering
  \subfigure[]{
  \includegraphics[clip=true,width=0.45\textwidth]{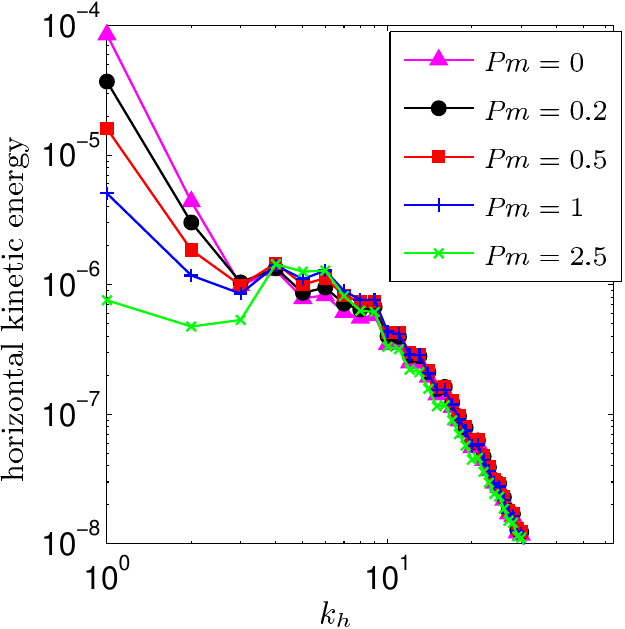}}
  \subfigure[]{
  \includegraphics[clip=true,width=0.45\textwidth]{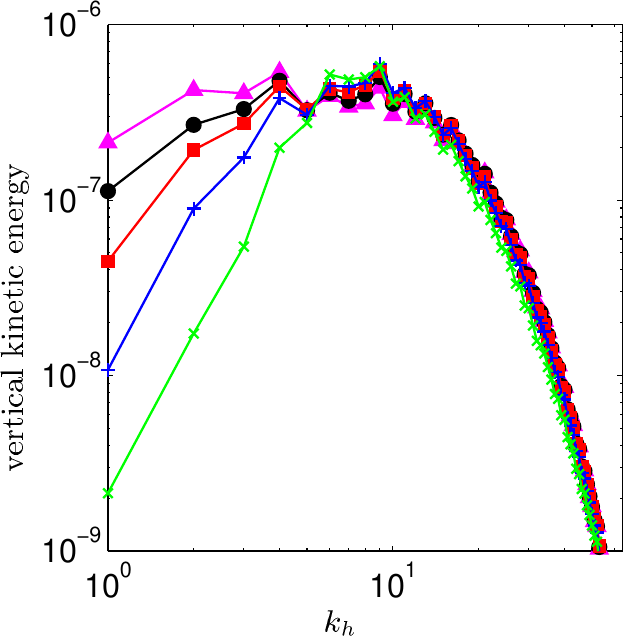}}
  \caption{
Time-averaged horizontal power spectra of (a) the horizontal kinetic energy $E_u^h$, and (b) the vertical 
  kinetic energy $E_u^v$, for the same cases as in figure~\ref{fig:specb_Ra5e8}, together with the hydrodynamical case ($\Pm=0$).}
\label{fig:specu_Ra5e8}
\end{figure}

Figure~\ref{fig:specu_Ra5e8} shows the time-averaged horizontal power spectra of the kinetic energies for $\Ra=5\times10^8$ and for different values of $\Pm$, including the comparison with the spectrum in the hydrodynamical case (denoted by $\Pm=0$). The spectrum for the horizontal flow at $\Pm=0.2$ shows that the velocity is dominated by the LSV (\ie $k_h=1$), in a similar fashion to the hydrodynamical case. As $\Pm$ is increased, the energy at large scales is progressively diminished, for both the horizontal and vertical flows. For example, the energy of the LSV is decreased by a factor $50$ for $\Pm=2.5$ compared with its value for $\Pm=0.2$. Figure~\ref{fig:wz_smalldyn} shows the axial vorticity for $\Pm=2.5$; the flow is dominated by a multitude of convective vortices, with no LSV visible. For sufficiently large $\Pm$ ($\Pm \gtrsim 1$ for $\Ra=5\times10^8$), the LSV is thus destroyed by the magnetic field. \citet{Guervilly2015} argued that it is the generation of the small-scale magnetic field that yields the suppression of the LSVs. We study this question in detail in \S\,\ref{sec:Suppression}. For small values of $\Pm$, a weaker small-scale magnetic field is generated, and consequently the LSV is at least partially preserved. Indeed, the presence of the LSV is vital to the generation of the large-scale magnetic field, as we shall see in \S\,\ref{sec:Mechanism}. 

\begin{figure}
\centering
  \subfigure[]{\label{fig:wz_l025}
  \raisebox{0cm}{\includegraphics[clip=true,width=0.4\textwidth]{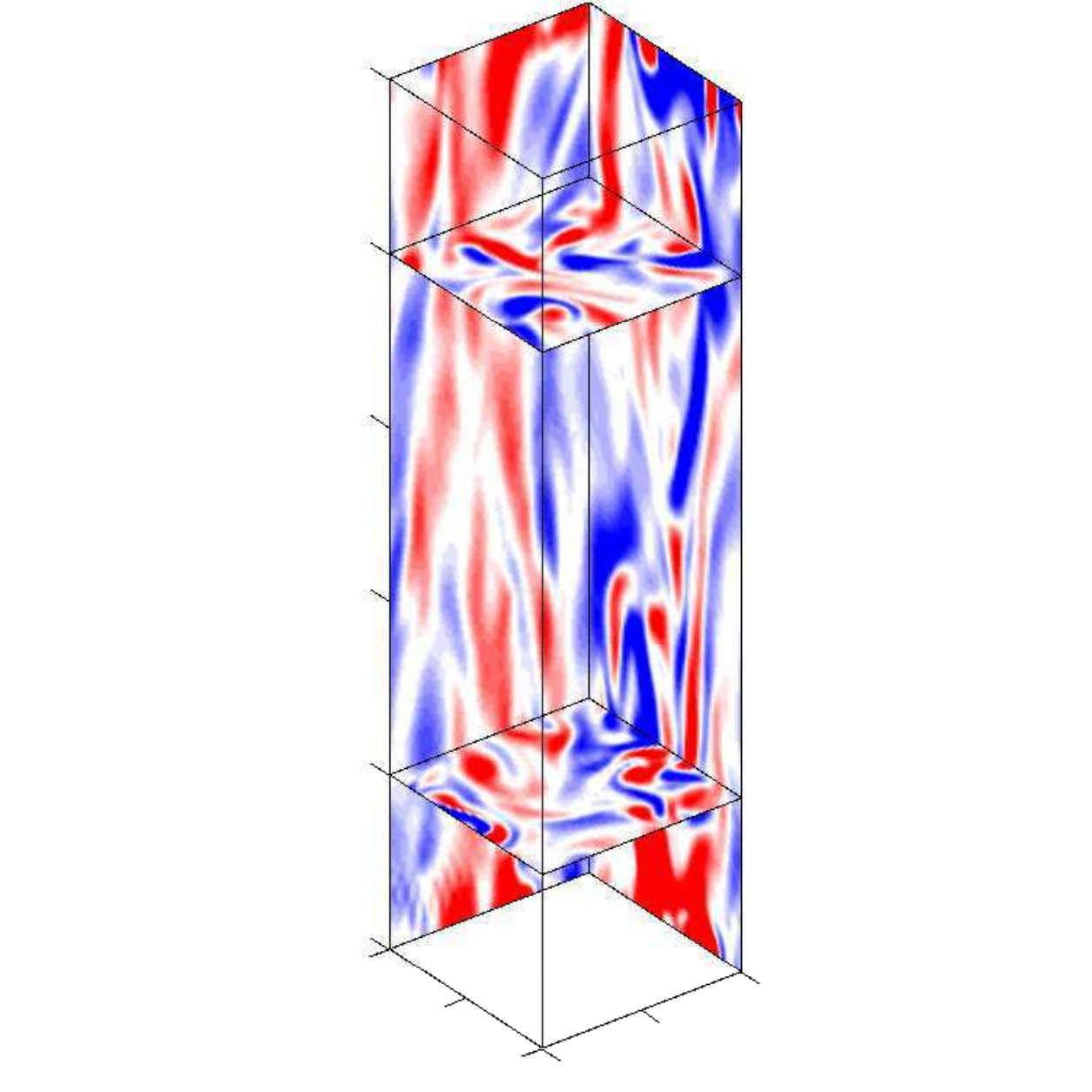}}}
  \subfigure[]{\label{fig:ME_smallscaledyn}
  \includegraphics[clip=true,width=0.48\textwidth]{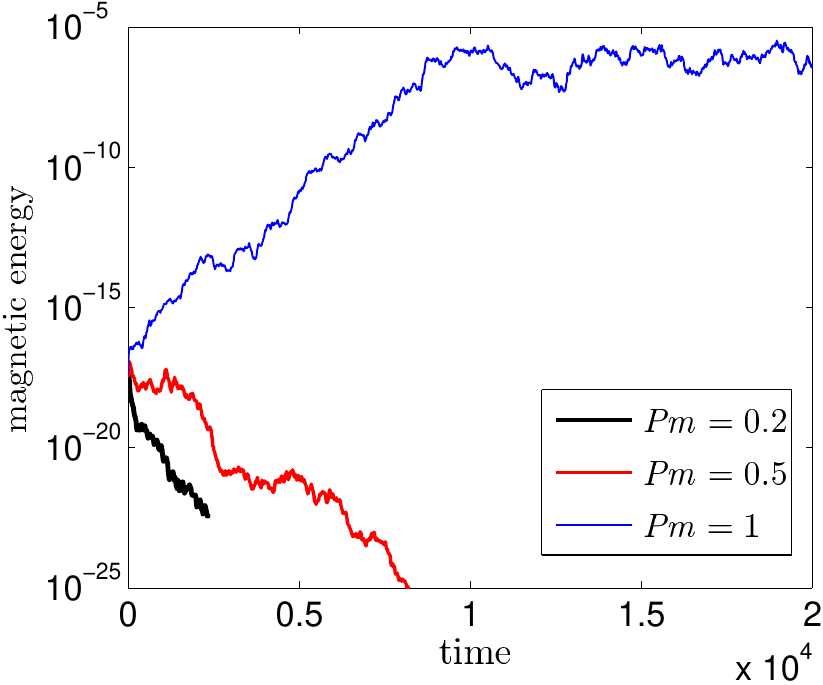}}
  \caption{
 (a) Horizontal and vertical cross-sections of axial vorticity for purely hydrodynamic convection with a small aspect ratio ($\aspect=0.25$) and $\Ra=5\times10^8$.
 (b) Time evolution of the magnetic energy for $\aspect=0.25$, $\Ra=5\times10^8$ and three values of $\Pm$.}
\end{figure}

To conclude this section, we determine the boundaries in parameter space of small-scale and LSV dynamo action. First we calculate the small-scale dynamo threshold. To do this, we perform simulations in a box of aspect ratio $\aspect=0.25$, chosen to be sufficiently small that no LSV forms, as can be seen from the horizontal cross-sections of the axial vorticity shown in figure~\ref{fig:wz_l025}. Figure~\ref{fig:ME_smallscaledyn} shows the time evolution of the magnetic energy for $\Ra=5\times10^8$ and three different values of $\Pm$ with this small aspect ratio. For $\Pm \lesssim 0.5$, the magnetic energy decays, whereas for $\Pm=1$, the magnetic energy increases and eventually saturates. The sustained magnetic field is dominated by small scales. The small-scale dynamo threshold is thus located between $\Pm=0.5$ and $\Pm = 1$. To complete the diagram of figure~\ref{fig:diagram}, we computed dynamo simulations with $\aspect=0.25$ for a range of $\Ra$. From these we find that the small-scale dynamo threshold is always in agreement with the dynamo transition identified earlier for $\aspect=1$, which was calculated by comparing the integral convective and magnetic wavenumbers (marked as diamonds in figure~\ref{fig:diagram}). The boundary between the two types of dynamo corresponds to the line where $\Rm \approx 550$; this is plotted in figure~\ref{fig:diagram}. 
It should though be noted that this line is indicative; the actual transition between the two types of dynamo is continuous.
We can also define a local magnetic Reynolds number, $\Rm_l = \Rm/k_u^v$. The curve $\Rm_l = 45$ lies close to that of $\Rm=550$ for all $\Ra$, since the integral convective wavenumber remains close to $k_u^v = 12$ for $\Ra$ increasing up to $\Ra = 10^9$. The small-scale dynamo threshold can therefore be defined equivalently by the local magnetic Reynolds number taking the value $\Rm_l \approx 45$. This threshold is independent of the formation of the LSV and, indeed, exists for $\Ra<3\times 10^8$, where no LSVs are present.

The lower threshold of the LSV dynamo is determined from consideration of kinematic dynamo action as $\Pm$ is decreased; this is shown as the red dashed line in figure~\ref{fig:diagram}. For $\Ra > 5 \times 10^8$, the threshold remains fixed at $\Pm \approx 0.15$, at least up to the highest Rayleigh number calculated, $\Ra=10^9$. For larger $\Ra$, there are two reasons why the LSV dynamo might fail: the LSVs themselves might cease to exist at large Rossby number (cf.\ figure~\ref{fig:diagram_LSV}), or the small-scale dynamo threshold intersects that of the LSV dynamo. From hydrodynamical simulations, the strength of the LSV declines when the local Rossby number, $\Ro_l$, exceeds a value of about $0.1$. For the Ekman number considered here, this occurs for $\Ra > 1.8 \times 10^9$ (\ie $\Ra / Ra_c > 18$). Since it was not possible to pursue dynamo simulations for $\Ra>10^9$, we were unable to locate directly the intersection of the small-scale and LSV dynamo thresholds. However, we could perform hydrodynamical simulations for larger $\Ra$, up to $\Ra=2.2\times10^9$ ($\Ra /\Ra_c \approx 22$), thereby allowing us to make a prediction about the demise of the LSV dynamo under the assumption that the small-scale dynamo threshold continues to obey $\Rm_l \approx 45$ for $\Pm<0.5$. We find that this boundary intersects the line $\Pm=0.15$ for some value of $\Ra$ in the range $2 \times 10^9 < \Ra < 2.2 \times 10^9$. For this value of $\Ek$, the two critical Rayleigh numbers --- one marking the demise of the LSV, the other the preference for small-scale dynamo action -- are, coincidentally, of the same order of magnitude.

\section{The LSV dynamo mechanism}
\label{sec:Mechanism}

LSVs, which consist predominantly of horizontal flows, do not of themselves act as a dynamo. There are therefore two possible mechanisms for the generation of large-scale magnetic field by an LSV dynamo. One is that the small-scale three-dimensional flows are modified crucially by the LSV in such a way that they are able to support dynamo action --- recall, from the results of figure~\ref{fig:ME_smallscaledyn}, that in this parameter regime, small-scale dynamo action without the influence of the LSV is not supported. The alternative is that LSV dynamo action arises through an essential combination of the LSV and the three-dimensional flows. In this section, we examine the elementary components of the LSV dynamo mechanism by considering filtered versions of the velocity field, as described in \citet{Hughes2013}. The idea is to study the kinematic dynamo problem for velocity fields obtained from the full convective velocity by the removal of selected spectral modes, with the aim of identifying which modes are the key players in the dynamo process. Note that only the velocity is filtered.
In order to understand the dynamo process, it is necessary to retain all modes in the description of the magnetic field; however, as such, this restricts attention to the kinematic dynamo problem.
The induction equation is then solved using the filtered velocity, but retaining all modes in the description of the magnetic field. 
The filtration of spectral modes is easily implemented in our code, but this type of filtration does not preserve the spatial coherence of the velocity structures. 
Nonetheless, it is a convenient and potentially instructive method of separating lengthscales in the flow.
In order to examine the importance of various modes in the velocity, we perform three different types of filtration: the removal of the LSV mode (LSV filtering), the removal of modes at high wavenumbers (short-wavelength filtering) and the removal of modes at low wavenumbers (long-wavelength filtering). We consider these separately in \S\S\,\ref{sec:lsv}, \ref{sec:swf}, \ref{sec:lwf} below, where we focus on the dynamo simulation with $\Ra=5\times10^8$ and $\Pm=0.3$; for this Rayleigh number, the Reynolds number is given by $\Rey=765$, and hence $\Rm=\Rey\Pm=230$.

\begin{figure}
\centering
  \subfigure[]{\label{fig:wz_filter_a}
  \includegraphics[clip=true,width=0.32\textwidth]{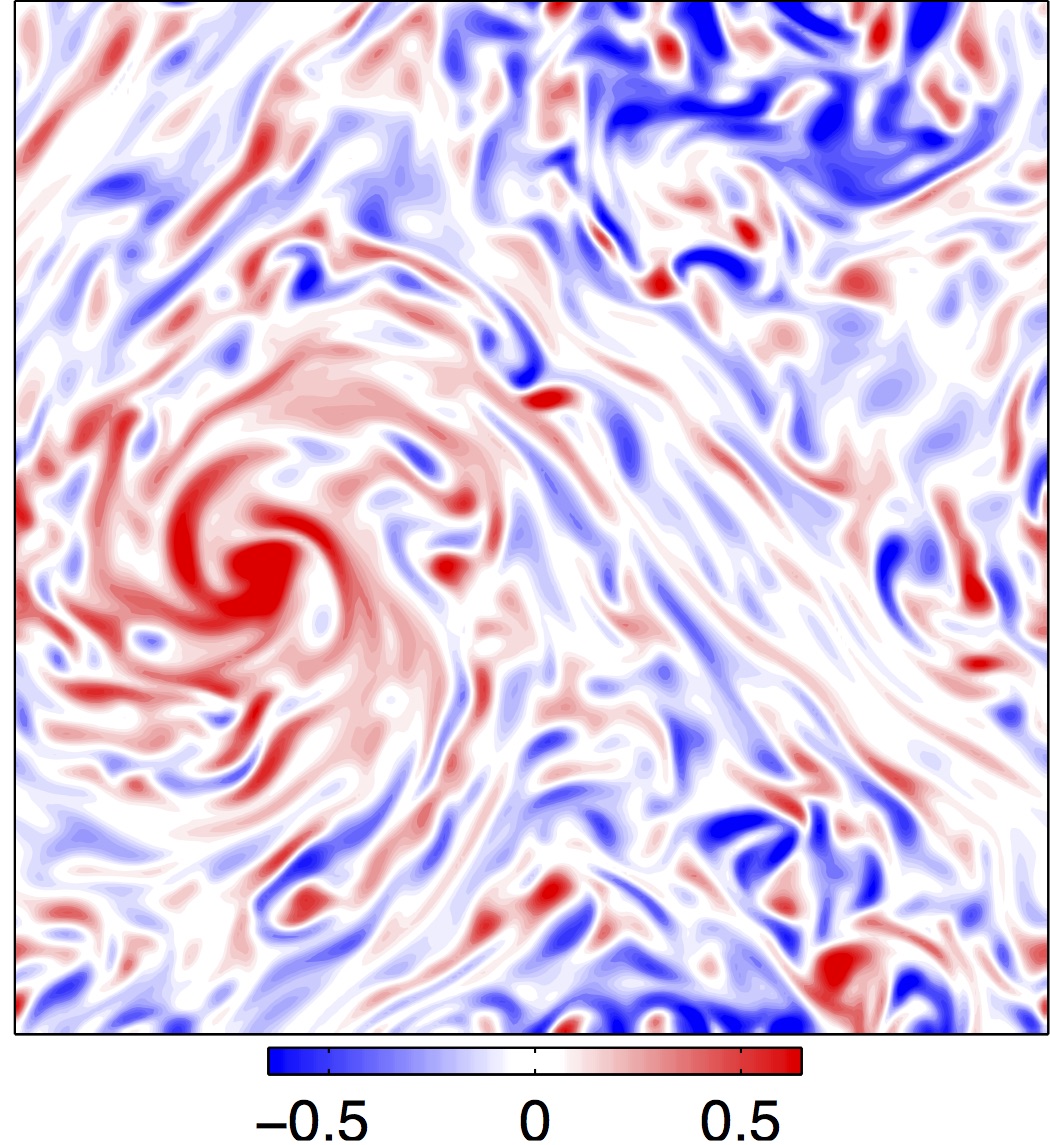}}
  \subfigure[]{\label{fig:wz_filter_b}
  \includegraphics[clip=true,width=0.32\textwidth]{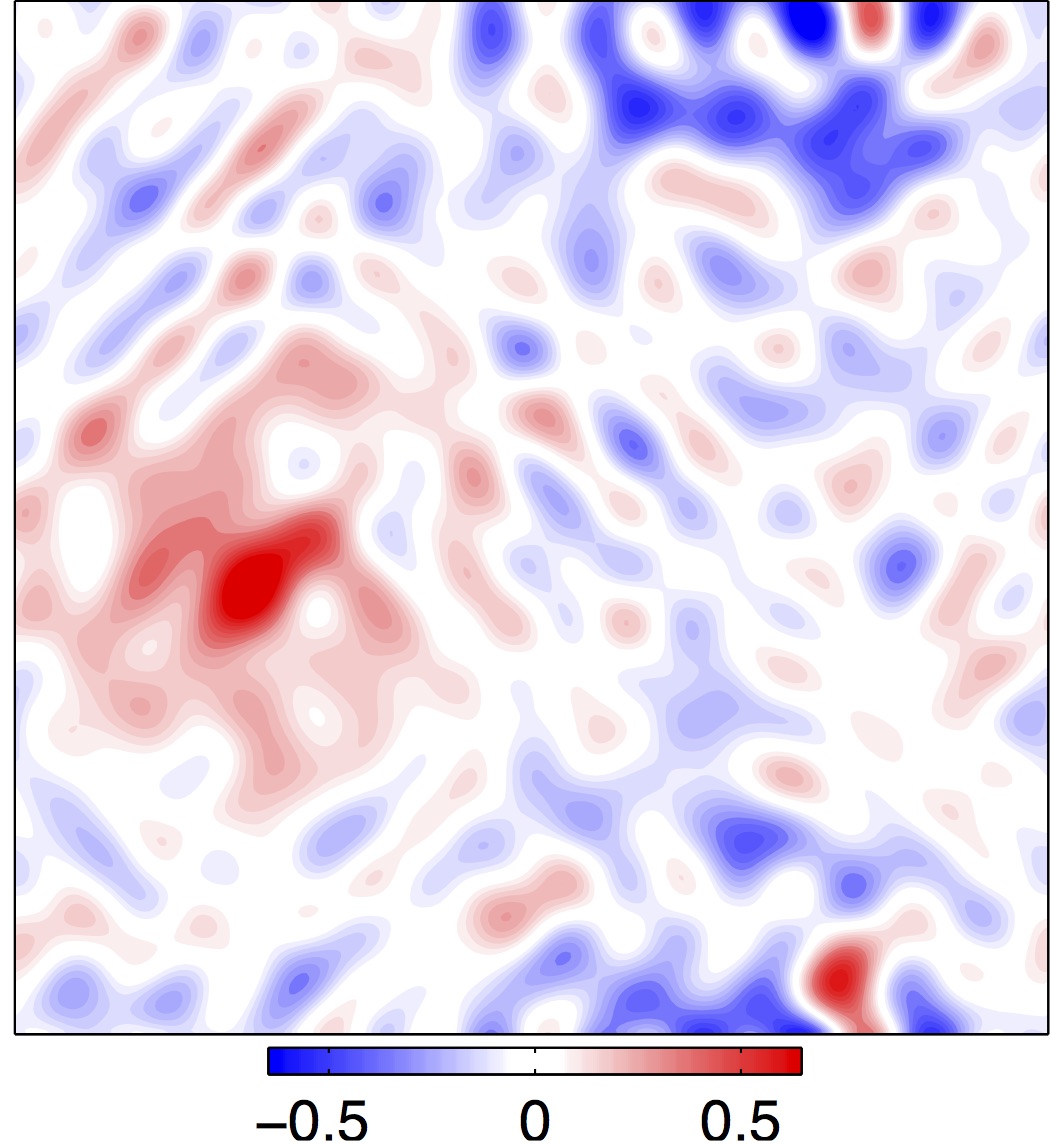}}
  \subfigure[]{\label{fig:wz_filter_c}
  \includegraphics[clip=true,width=0.32\textwidth]{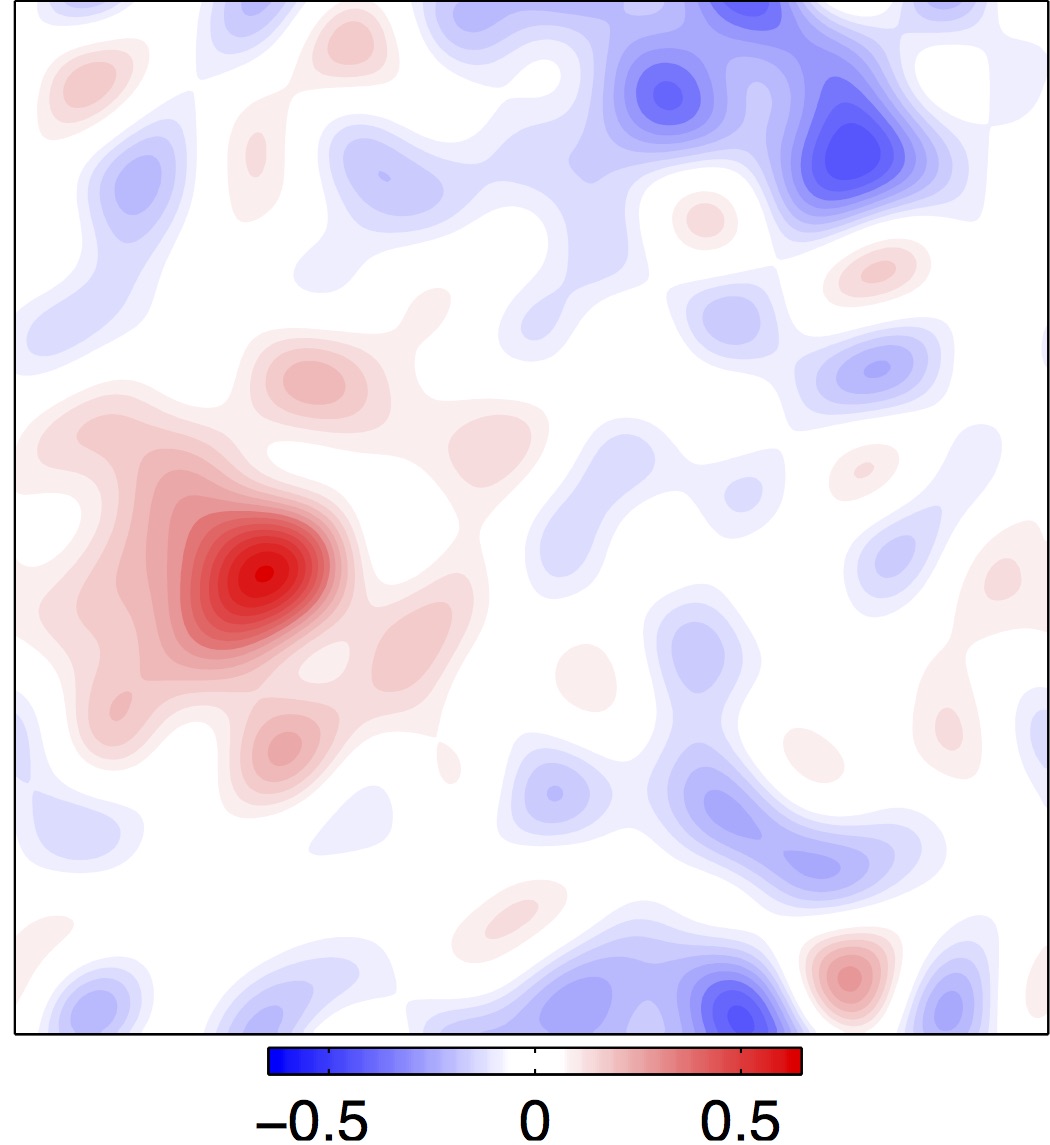}}
  \subfigure[]{\label{fig:wz_filter_f}
  \includegraphics[clip=true,width=0.32\textwidth]{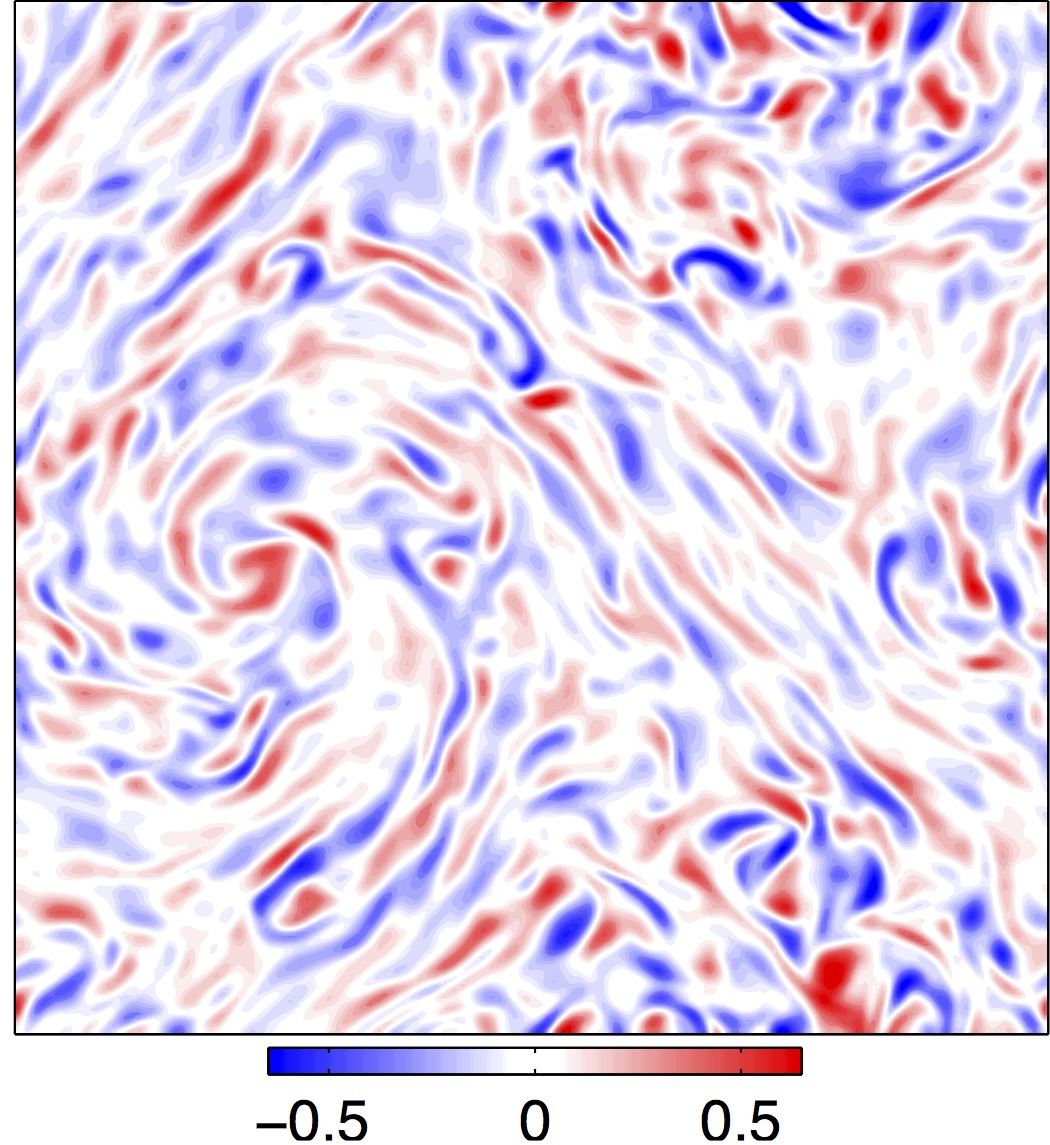}}
  \subfigure[]{\label{fig:wz_filter_d}
  \includegraphics[clip=true,width=0.32\textwidth]{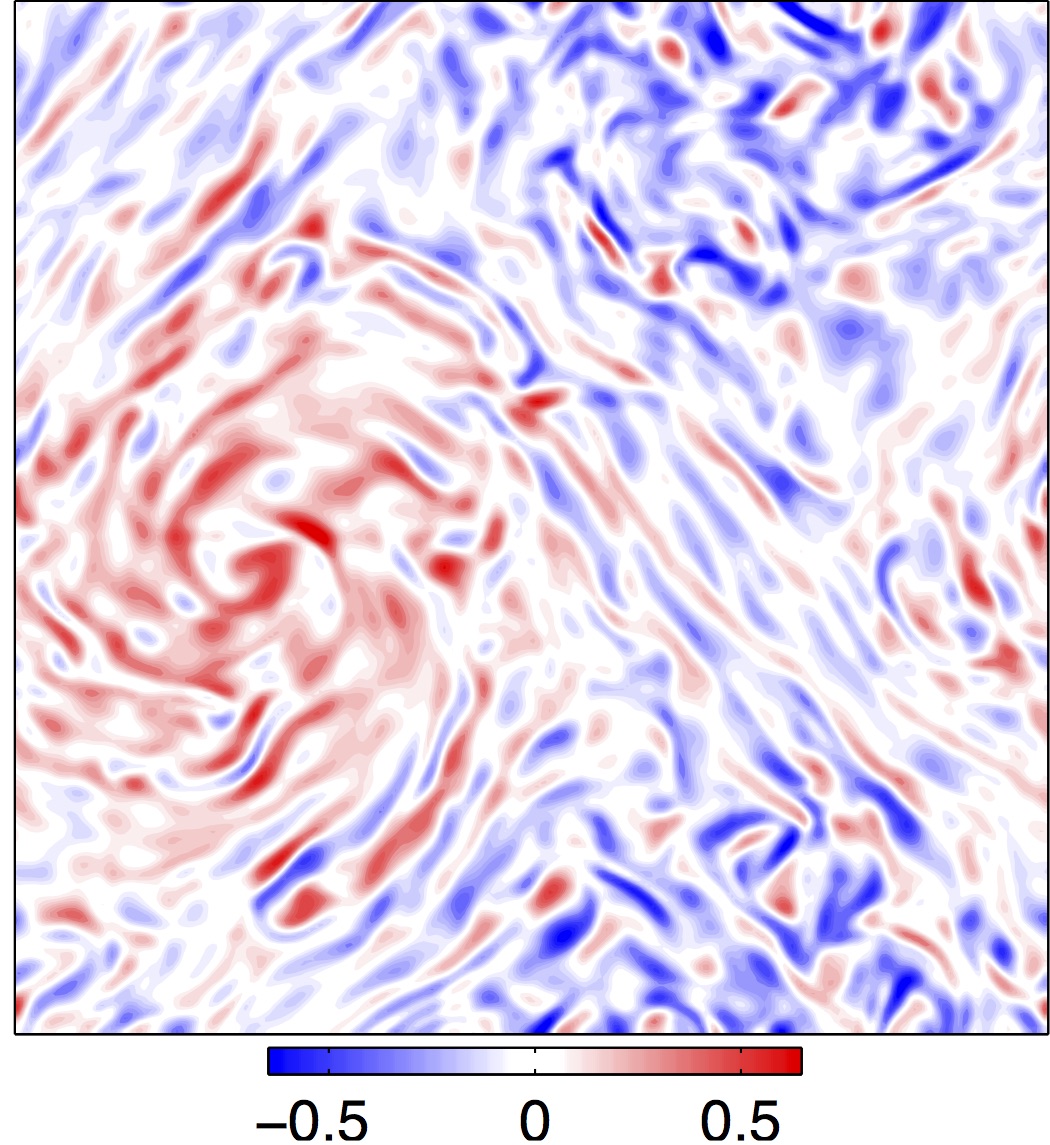}}
  \subfigure[]{\label{fig:wz_filter_e}
  \includegraphics[clip=true,width=0.32\textwidth]{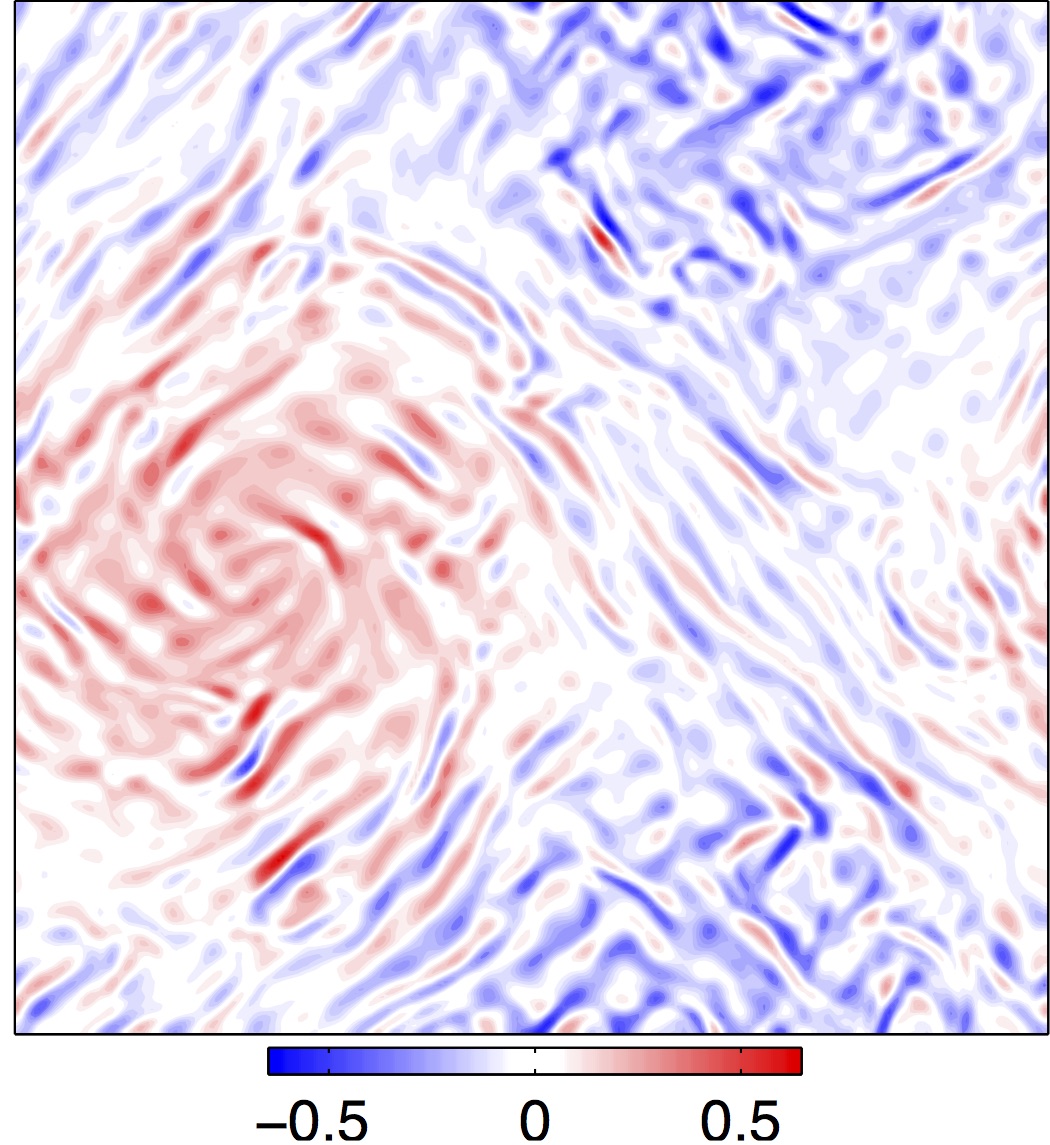}}
  \caption{
  Snapshots of the axial vorticity of the flow used in the filtered simulations ($\Ra=5\times10^8$, $\Pm=0.3$): 
  (a) unfiltered flow; short-wavelength cutoff: (b) $k_f=10$ and (c) $k_f=6$; (d) LSV-filtered flow;
  long-wavelength cutoff: (e) $k_f=4$ and (f) $k_f=8$.}
\label{fig:wz_filter}
\end{figure}

\subsection{LSV filtering}\label{sec:lsv}

We first examine the role played by the LSV in the dynamo process: in particular, does the LSV velocity field contribute directly to the magnetic induction, or does it play a more subtle role through modifying the smaller-scale flows in such a way that they become favourable to dynamo action? To address this question, we consider the velocity field obtained by removing only the modes \mbox{$(k_x,k_y) \leq (1,1)$} from the velocity (LSV filtering). A snapshot of this filtered flow is shown in figure~\ref{fig:wz_filter_f}, illustrating that the effect of the LSV on the rest of the flow is kept intact; for comparison, the unfiltered flow is shown in figure~\ref{fig:wz_filter_a}. The result of the kinematic dynamo simulation is shown in figure~\ref{fig:ME_filter_b} (labelled LSV-filter), leading to the interesting result that the filtered flow fails to act as a dynamo. Consequently, despite its inability to sustain a dynamo by itself, the LSV must play a critical role in the magnetic induction; in particular, its importance for dynamo action is not confined to its hydrodynamical influence on the other parts of the flow.

\subsection{Short-wavelength filtering}\label{sec:swf}

\begin{figure}
\centering
  \subfigure[]{\label{fig:ME_filter_a}
  \includegraphics[clip=true,width=0.48\textwidth]{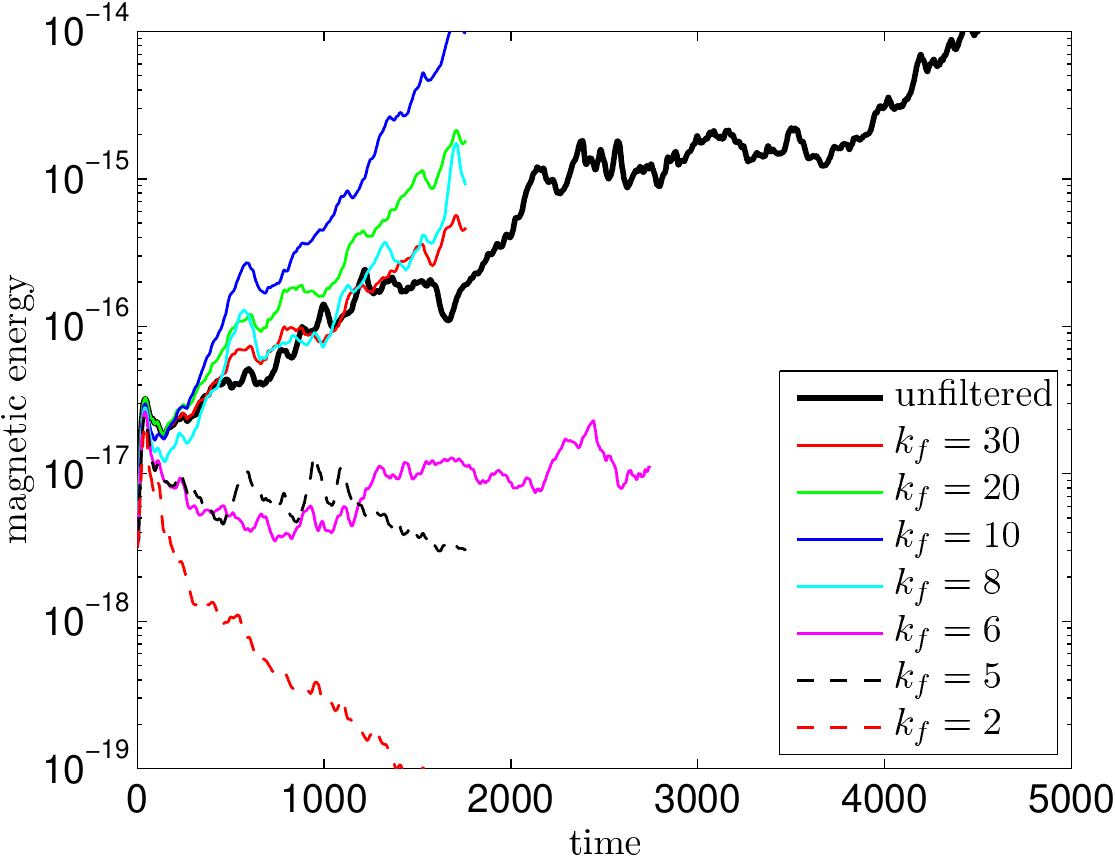}}
  \subfigure[]{\label{fig:ME_filter_b}
  \includegraphics[clip=true,width=0.48\textwidth]{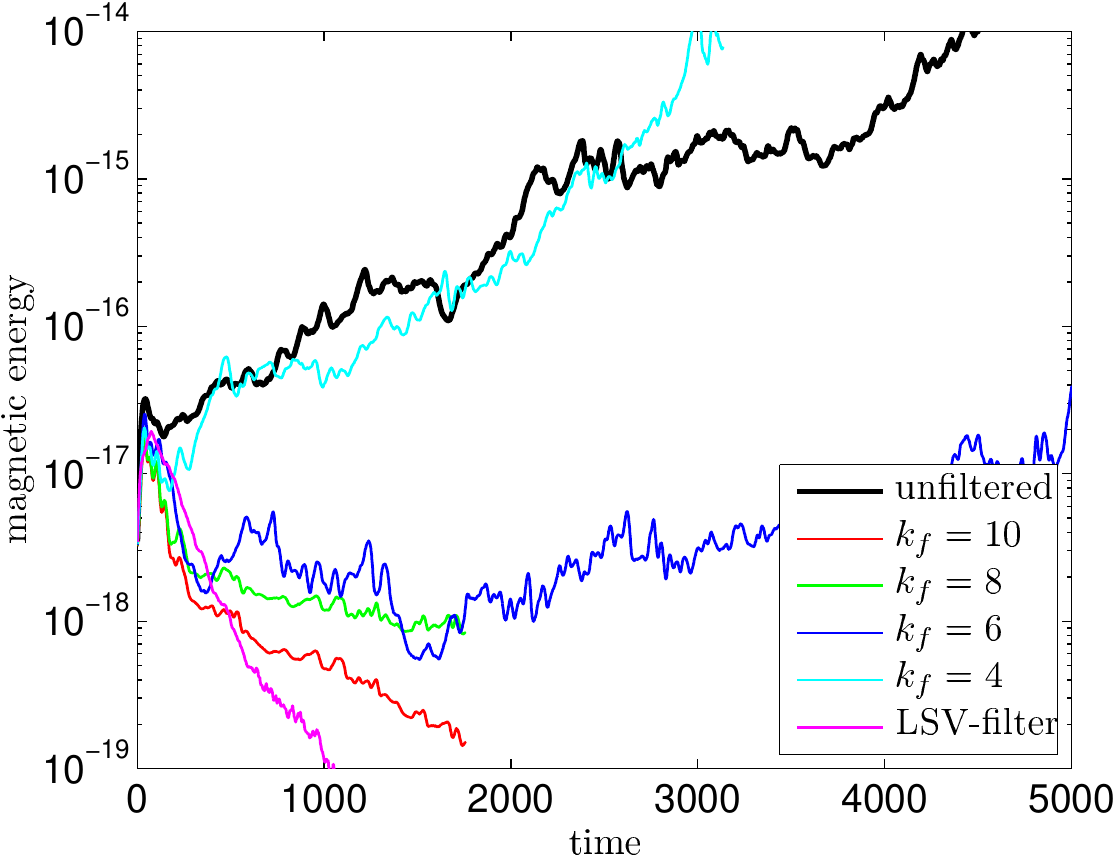}}
  \caption{Time series of the magnetic energy of the filtered simulations ($\Ra=5\times10^8$, $\Pm=0.3$):
  (a) short-wavelength cutoff; (b) long-wavelength cutoff and LSV filtration.}
\end{figure}

We now consider the nature of the dynamo action that arises from short-wavelength filtration of the velocity field. 
The cutoff horizontal wavenumber is denoted by $k_f$: all the spectral modes for which either \mbox{$k_x > k_f$} or \mbox{$k_y>k_f$} are set to zero. 
Examples of short-wavelength filtered flows are shown in figures~\ref{fig:wz_filter_b}-\ref{fig:wz_filter_c}. Figure~\ref{fig:ME_filter_a} shows time series of the magnetic energy from simulations in which the cutoff wavenumber is varied from $k_f = 30$ to $k_f = 2$; the magnetic energy of the kinematic dynamo of the unfiltered flow is also plotted for comparison. As $k_f$ is decreased from the unfiltered flow to $k_f = 10$, the dynamo is successfully maintained by the filtered flows; indeed, the growth rate of the magnetic energy increases, indicating that the small scales are simply providing enhanced dissipation. For $k_f=8$, the flow still acts as a dynamo, but with a smaller growth rate than that of $k_f=10$, thereby indicating that the filtration has started to impede the dynamo mechanism. For $k_f = 6$, the magnetic energy is essentially neither decreasing nor increasing, \ie this case is close to the dynamo threshold. For $k_f  \le 5$, the dynamo fails. The intermediate modes of scale between that of the box size and that given by $k_h = O(8-10)$ are therefore absolutely vital for dynamo action at $\Pm=0.3$. Visual comparison of a flow that is capable of dynamo action (figure~\ref{fig:wz_filter_b} for $k_f=10$) with one that is not (figure~\ref{fig:wz_filter_c} for $k_f=6$), suggests that the  elongated flow structures that are produced by the horizontal shear from the LSV might be crucial for the dynamo process.

\begin{figure}
\centering
  \subfigure[]{\label{fig:B_filter_a}
  \includegraphics[clip=true,width=0.32\textwidth]{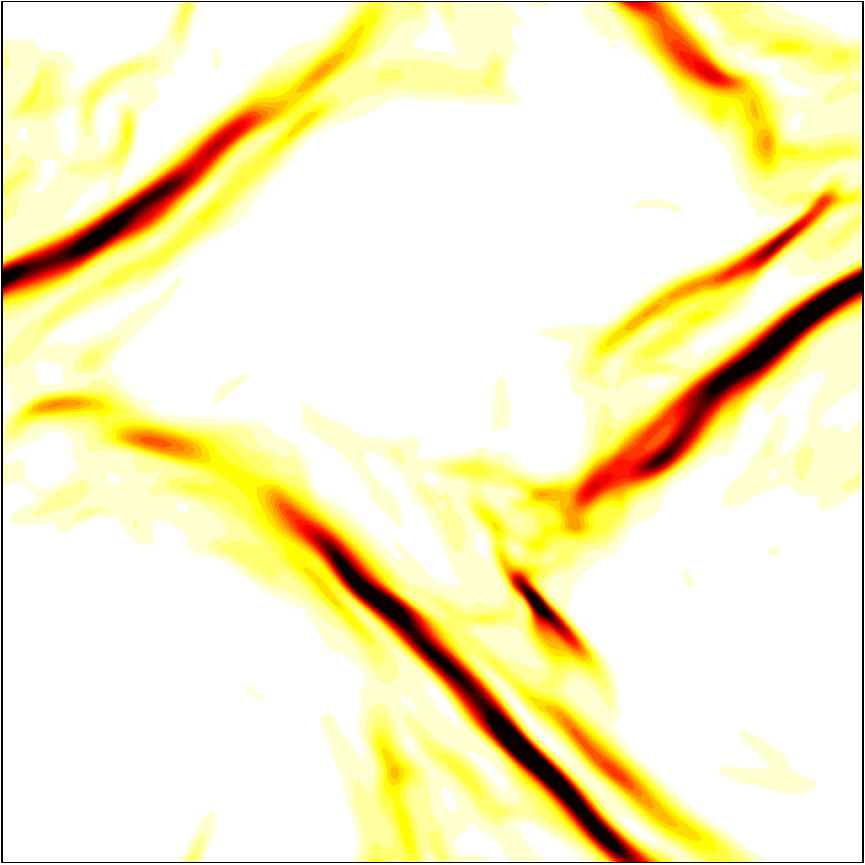}}
  \subfigure[]{\label{fig:B_filter_b}
  \includegraphics[clip=true,width=0.32\textwidth]{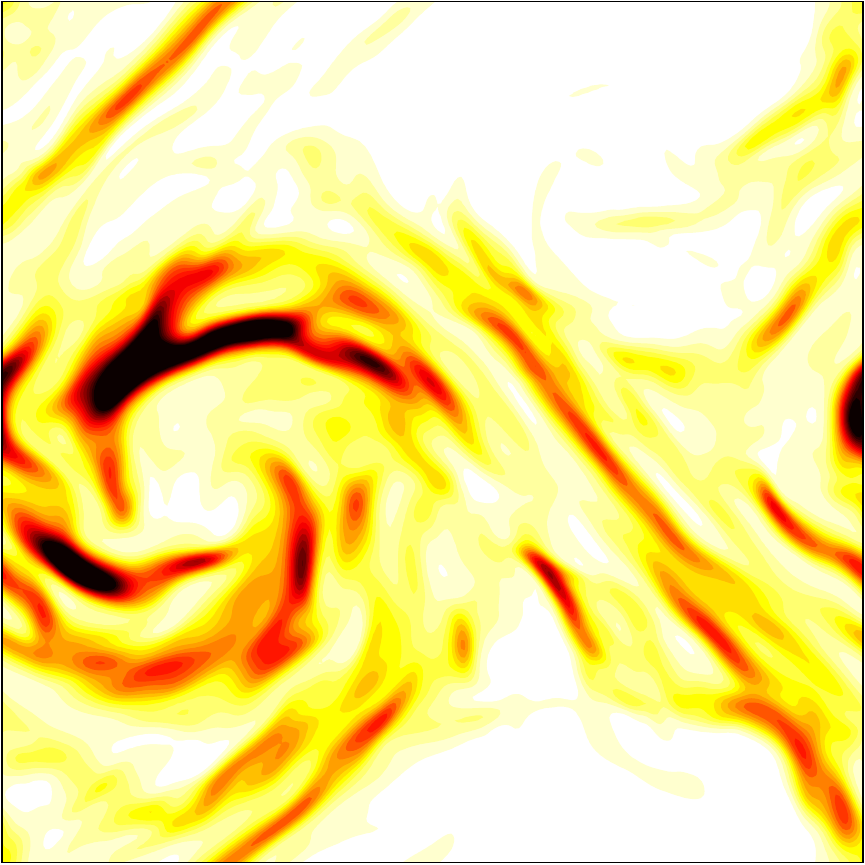}}
  \subfigure[]{\label{fig:B_filter_c}
  \includegraphics[clip=true,width=0.32\textwidth]{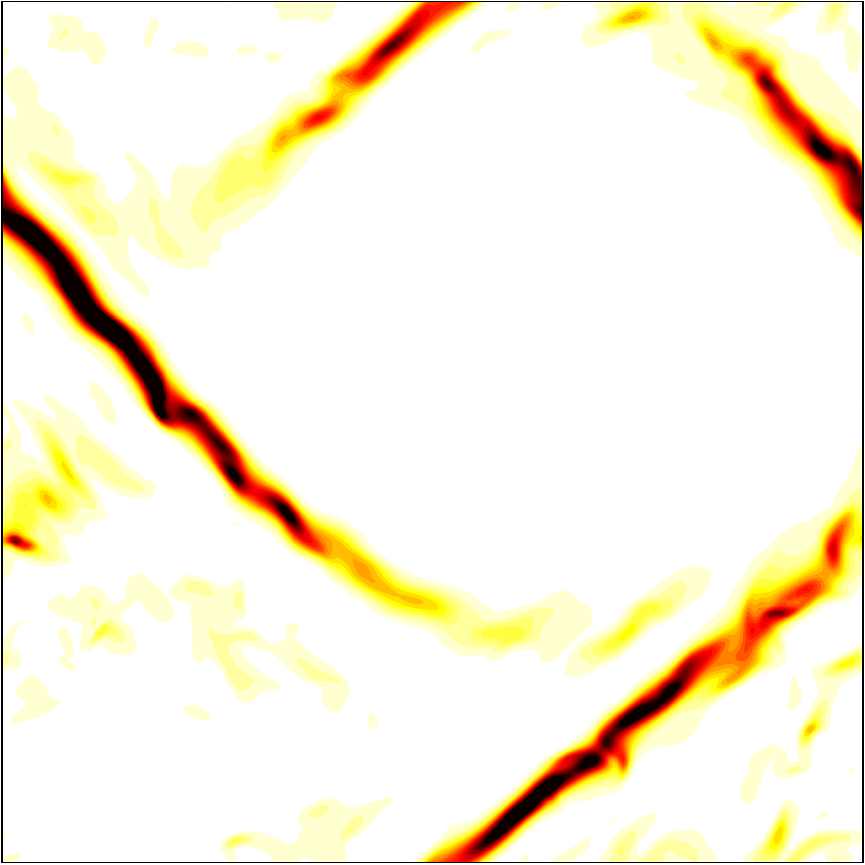}}
  \caption{Snapshots of \mbox{$\langle B_x^2 + B_y^2 \rangle_z$} for (a)~the unfiltered flow (kinematic dynamo), and flows with (b)~short-wavelength cutoff ($k_f=10$), and (c)~long-wavelength cutoff ($k_f=4$).}
 \label{fig:B_filter} 
\end{figure}

Figure~\ref{fig:B_filter_b} shows the structure of the horizontal magnetic field for the filtered simulation with $k_f = 10$; for comparison, the field produced by the unfiltered flow in the kinematic case is shown in figure~\ref{fig:B_filter_a}. The plotted quantity is a snapshot of \mbox{$\langle B_x^2 + B_y^2 \rangle_z$}, twice the depth-averaged horizontal magnetic energy. As in the unfiltered case, the horizontal magnetic field appears on a large scale and is organised into bands around the cyclone. However, unlike the unfiltered case, the horizontal field is also strong in the core of the  cyclone. This observation indicates that magnetic induction does occur in the core of the cyclone, but that the small-scale components of the unfiltered flow (\ie $k_h > k_f$) must subsequently expel the magnetic field from the core into the shear layer.

\begin{figure}
\centering
  \subfigure[]{
  \includegraphics[clip=true,width=0.23\textwidth]{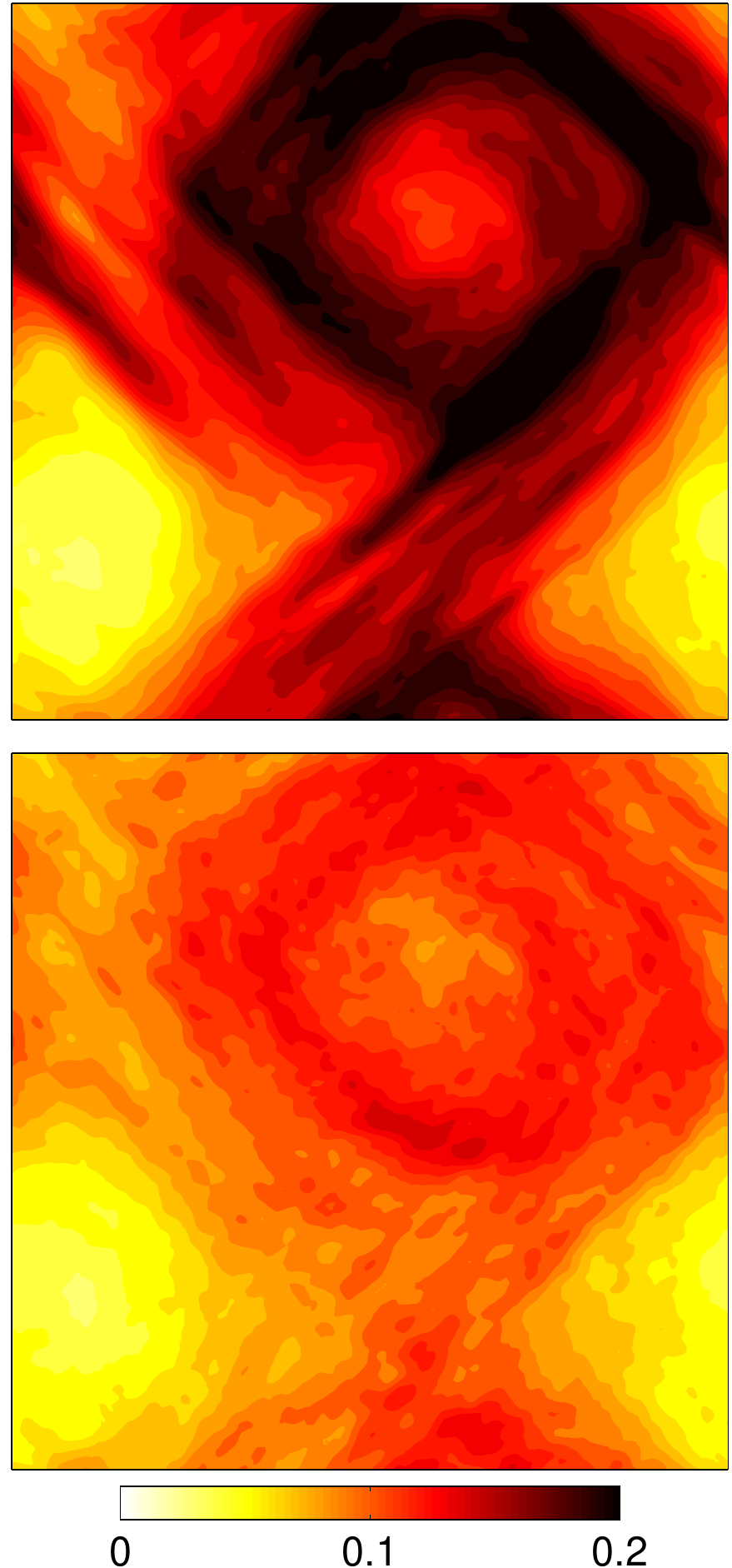}}
  \subfigure[]{
  \includegraphics[clip=true,width=0.23\textwidth]{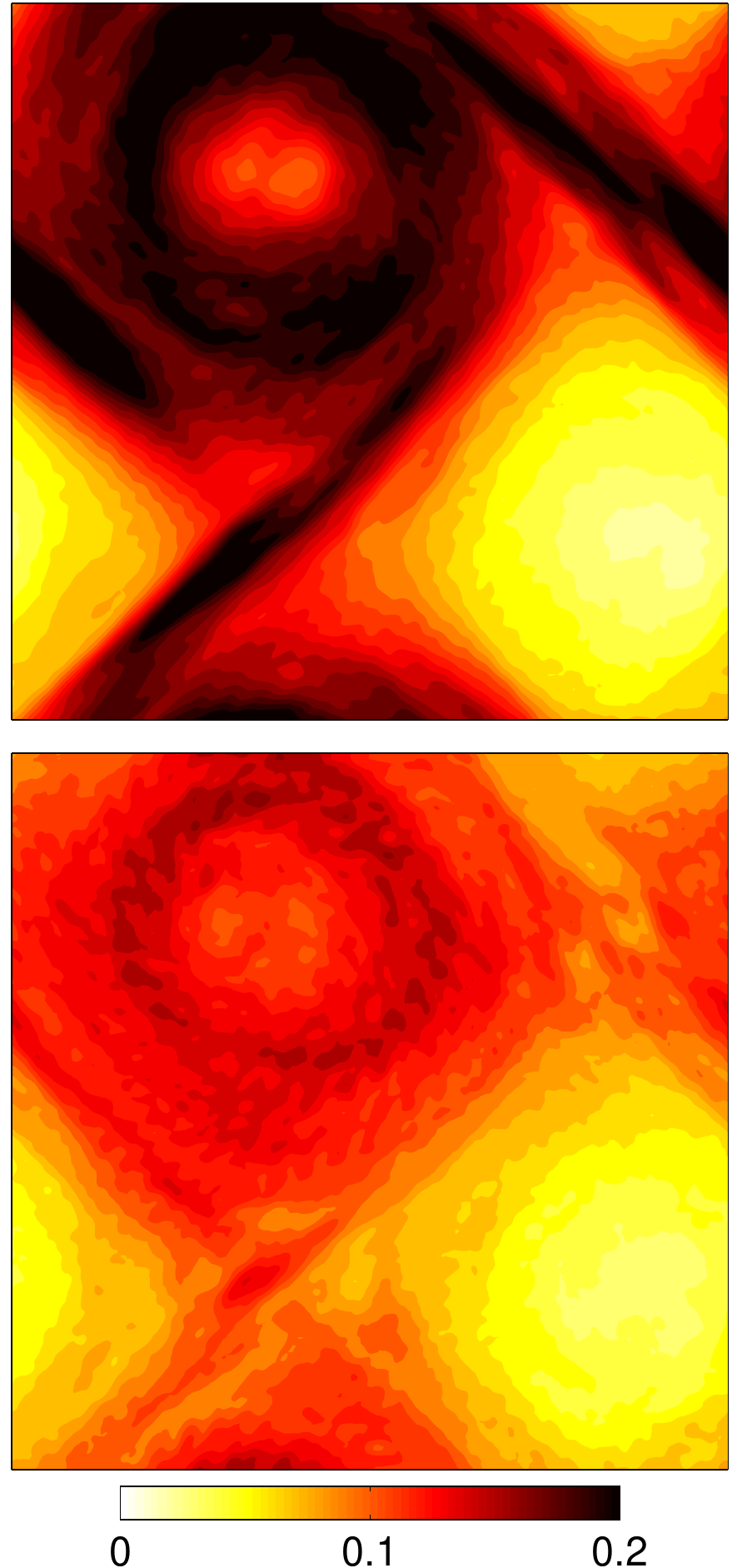}}
  \subfigure[]{
  \includegraphics[clip=true,width=0.23\textwidth]{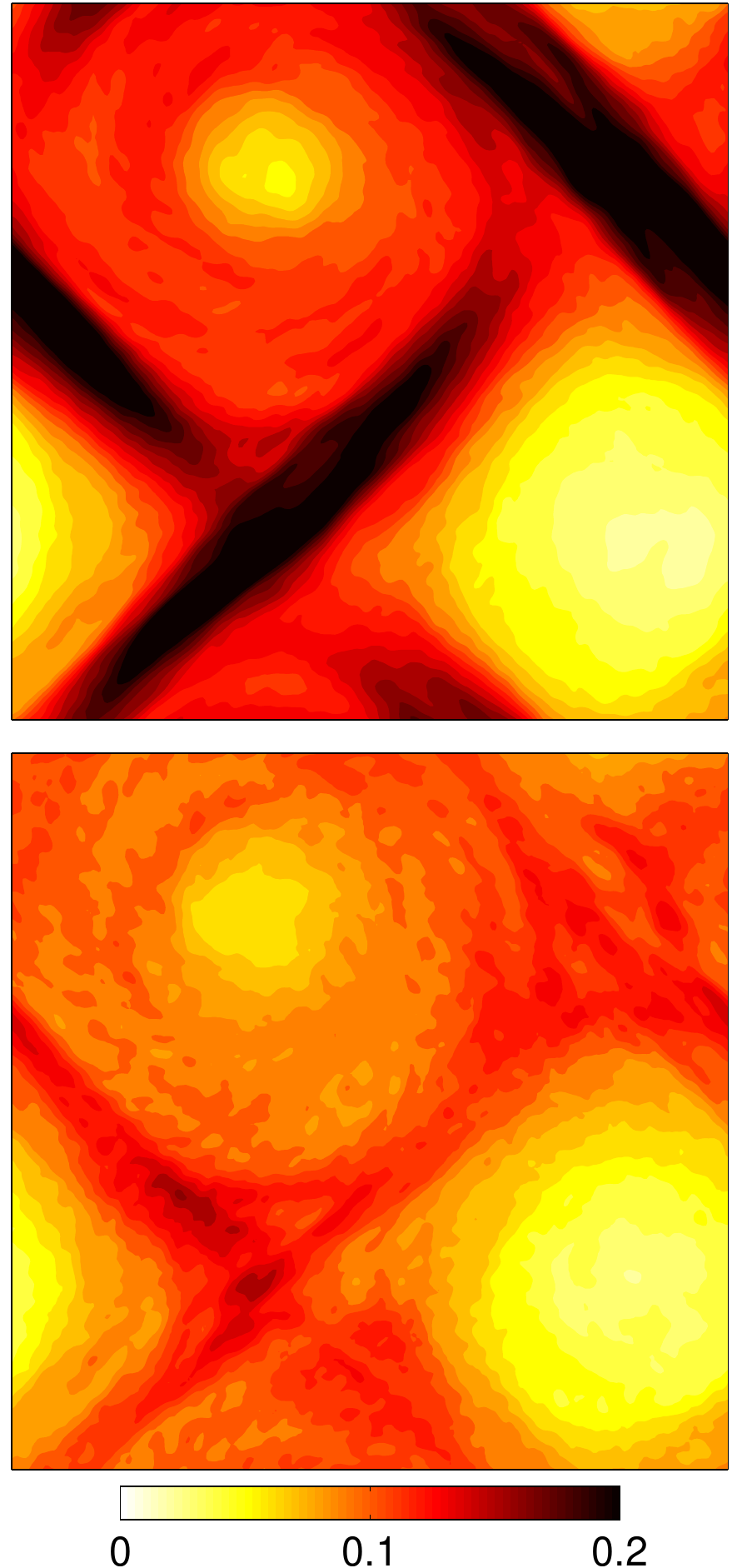}}
  \subfigure[]{
  \includegraphics[clip=true,width=0.23\textwidth]{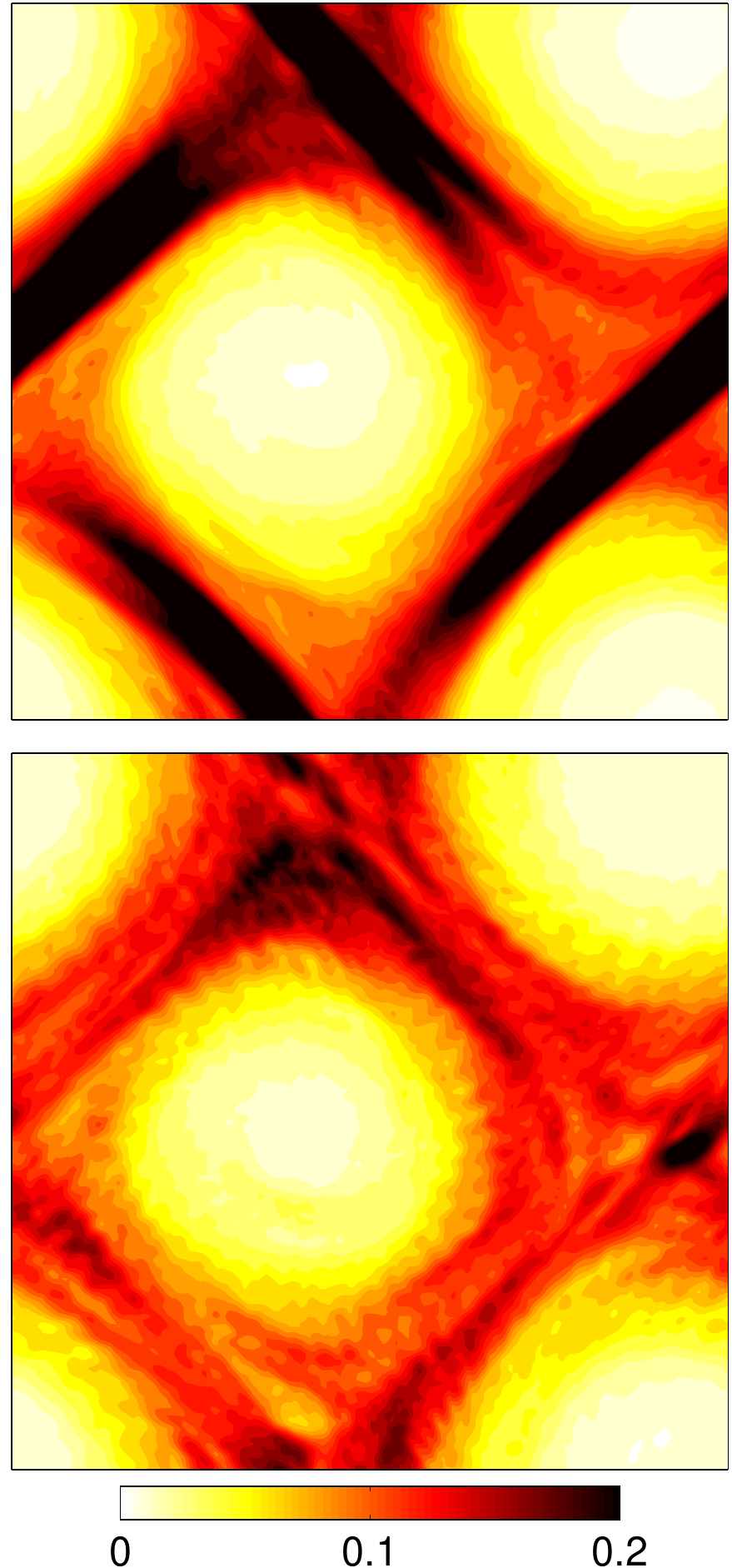}}
  \caption{Time-average of \mbox{$(B_x^2 + B_y^2)/\textrm{max}(B_x^2 + B_y^2)$} (top row)
  and \mbox{$B_z^2 /\textrm{max}(B_z^2)$} (bottom) in a horizontal slice ($z=0.25$) 
  for short-wavelength filtered dynamo simulations:
  (a) $k_f=10$; (b) $k_f=30$, (c) $k_f=40$; (d) unfiltered flow (kinematic dynamo).}
  \label{fig:Bave}
\end{figure}

The destruction of the field inside the core of the cyclone is portrayed in figure~\ref{fig:Bave}, which shows the time-averaged values of \mbox{$B_x^2 + B_y^2$} and \mbox{$B_z^2$} in a given horizontal slice, for different filtered simulations as well as for the unfiltered case. Since here we are considering kinematic dynamos, the values of \mbox{$B_x^2 + B_y^2$} and \mbox{$B_z^2$} have been normalised by their maximum value at each timestep before averaging. The time averages are taken over $400$ snapshots, during which the LSV drifts slightly in the horizontal plane; this explains why the averaged values are not as sharp as in the snapshot of figure~\ref{fig:B_filter_b}. For $k_f\leq30$, both the horizontal and vertical magnetic fields are concentrated mainly in the cyclone. As the filtration is relaxed ($k_f\geq40$), the horizontal and vertical magnetic fields are expelled from the core of the cyclone. It is therefore the flow corresponding to large wavenumbers ($k_h>30$) that leads to the magnetic flux expulsion. The dilute large-scale anticyclone always corresponds to a region of weak magnetic field; here, the horizontal shear is weaker than in the concentrated large-scale cyclone and seems unable to promote field amplification.

\subsection{Long-wavelength filtering}\label{sec:lwf}

In performing long-wavelength filtration, the amplitudes of the modes for which either \mbox{$k_x < k_f$} or \mbox{$k_y<k_f$} are set to zero, except, crucially, for the modes corresponding to the large-scale flow, \ie \mbox{$(k_x,k_y) \leq (1,1)$}. We have seen from \S\,\ref{sec:lsv} that the LSV plays a critical role in the magnetic induction. The aim here therefore is to determine whether a large-scale dynamo of  `$\alpha \omega$' type is at work, \ie a two-scale dynamo driven by a combination of the shearing at large scale (the $\omega$-effect) and the production of a coherent emf by interactions of the small scales (the $\alpha$-effect). Examples of flows obtained from a long wavelength filtration are shown in figures~\ref{fig:wz_filter_d}-\ref{fig:wz_filter_e}. Figure~\ref{fig:ME_filter_b} shows the temporal evolution of the magnetic energy from a series of filtered simulations in which the cutoff wavenumber is varied from $k_f=10$ to $k_f=4$. For $k_f \gtrsim 6$ the dynamo fails, but with $k_f=4$ the flow is able to sustain a magnetic field. These experiments thus demonstrate that the dynamo requires the presence of intermediate scales up to $k_h=4$, i.e.\ a scale somewhat larger than a characteristic convective scale (see figure~\ref{fig:scales_Ra5e8}), but that larger scales (with the exception of $k_h=1$) are not necessary for dynamo action at $\Pm=0.3$. There is therefore no rigorous scale separation between that of the shear exerted by the LSV and that of the modes that are significant for dynamo action ($k_f\approx 4- 10$); as such, this dynamo cannot strictly be categorised as an $\alpha \omega$-dynamo. Figure~\ref{fig:B_filter_c} shows the structure of the horizontal magnetic field for the case $k_f=4$. The field is similar to that produced in the unfiltered case, with bands of strong magnetic intensity concentrated in the shear layers surrounding the LSV, and with no magnetic field in the core of the cyclone.

In summary, the filtering exercise helps us to identify that the LSV dynamo mechanism (at least in the kinematic regime) relies on 
the presence of the LSV together with velocity modes of scale intermediate between the box size and the dominant convective scale.
The LSV itself plays a crucial role in the magnetic induction, and not simply via its action on the smaller-scale flows. 
The filtered simulations also show that the magnetic field is generated in the core of the LSV and in the surrounding shear layers. The magnetic field is subsequently 
expelled from the core by small-scale vortices.

\section{Suppression and Oscillations of the LSV}

\label{sec:Suppression}

In \S\,\ref{sec:TTD} we demonstrated the existence of a transition between an LSV dynamo and a small-scale dynamo as $\Pm$ is increased for a given convective flow. In this section we look in detail at the physical processes underlying this marked change in the nature of the dynamo, and in particular at how the LSV can be suppressed by the small-scale magnetic field. In \S\S\,\ref{sec:TTD}, \ref{sec:Mechanism} we focused attention chiefly on the spatial characteristics of the generated magnetic fields. Here, in order to gain an understanding of the transition between LSV and small-scale dynamos, we first consider the detailed temporal evolution of dynamos at different $\Pm$. In \S\,\ref{sec:5.1} we consider the case of $\Pm=2.5$, for which the LSV is destroyed, and for which dynamo action is unambiguously small-scale. In \S\,\ref{sec:5.2} we consider the same convective flow but with $\Pm=0.2$, which gives rise to an LSV dynamo. However, even in this case, the LSV feels the influence of the small-scale magnetic field, leading to temporal variations in the strength of the LSV. Then, in \S\,\ref{sec:5.3}, we explore the underlying physical processes by which the magnetic field can influence, or even destroy, the LSV.

\subsection{Suppression of the LSV}
\label{sec:5.1}

\begin{figure}
\centering
  \subfigure[]{\label{fig:oscillations_Ra5e8_Pm25}
  \includegraphics[clip=true,width=13cm]{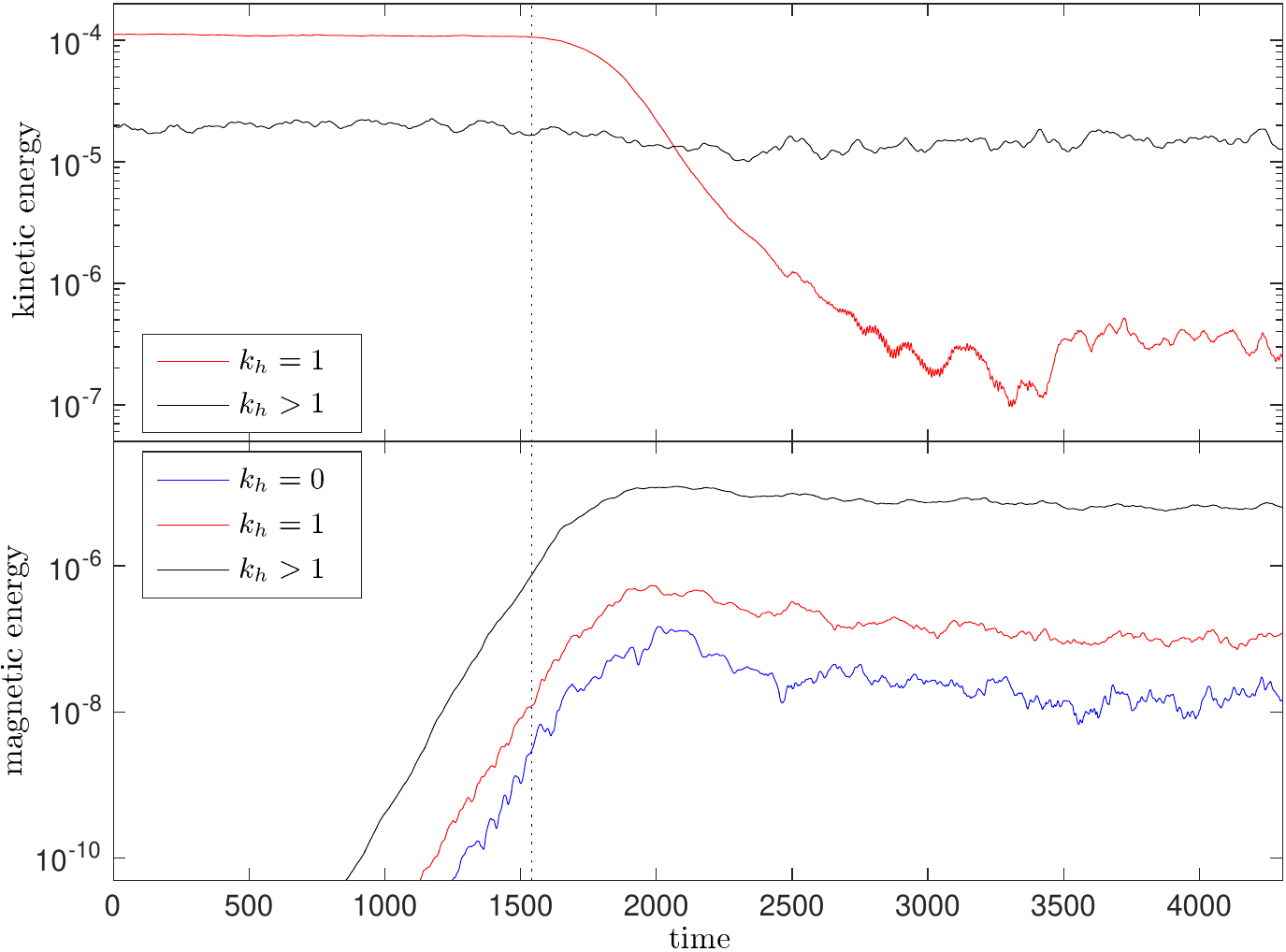}}
  \subfigure[]{\label{fig:oscillations_Ra5e8_Pm02}
  \includegraphics[clip=true,width=13cm]{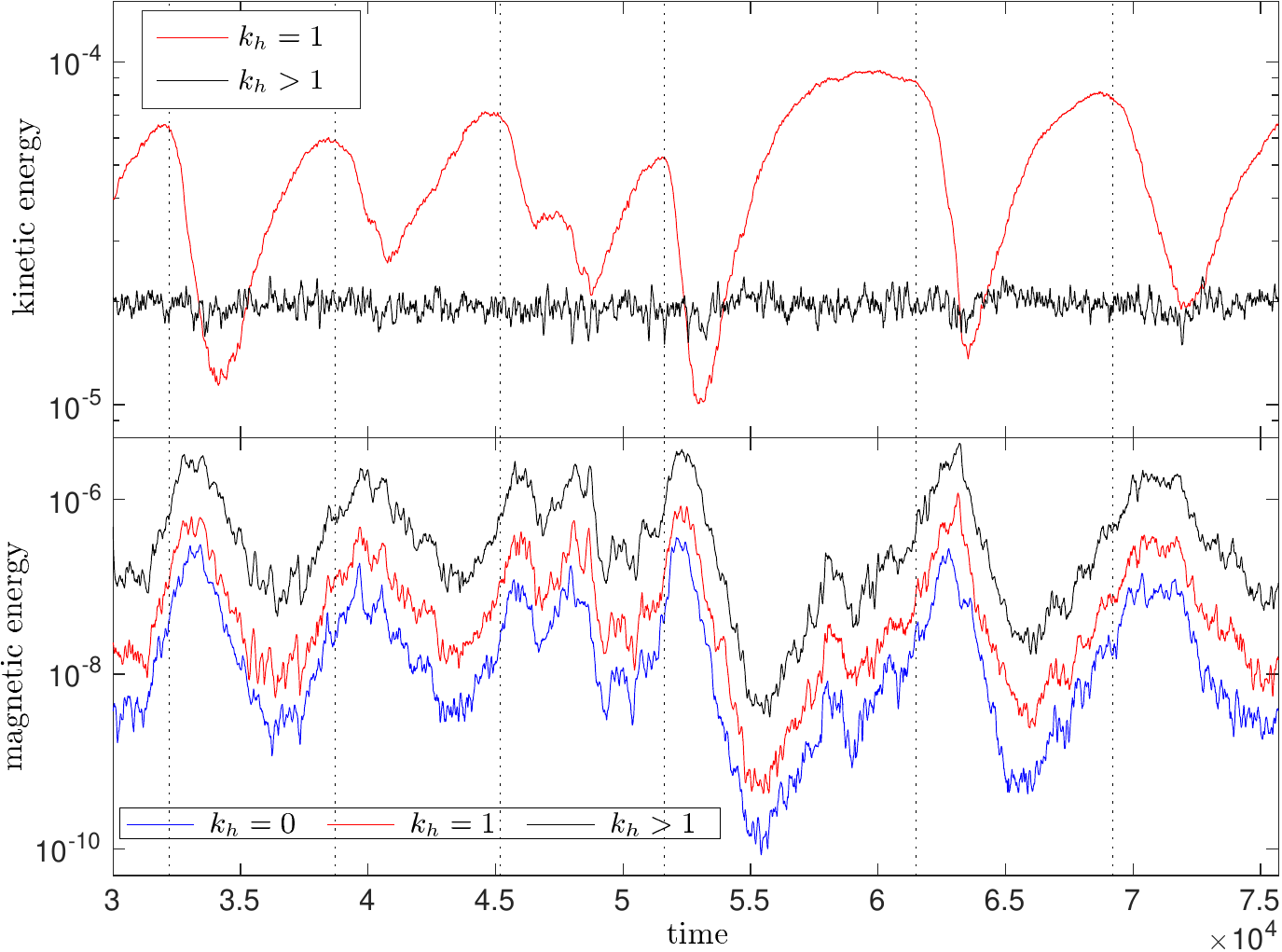}}
  \caption{Time series of the kinetic and magnetic energies of the modes $k_h=0$, $k_h=1$ and $k_h>1$
for the dynamos at $\Ra=5\times10^8$, (a) $\Pm=2.5$ and (b) $\Pm=0.2$.
The dotted vertical lines indicate the approximate times at which the kinetic energy of the LSV starts to decrease.
}
\label{fig:oscillations_Ra5e8}
\end{figure}

In this subsection we consider the dynamo resulting from the convective flow with $\Ra = 5 \times 10^8$ and $\Pm=2.5$; as can be seen from figure~\ref{fig:diagram}, it lies well within the small-scale dynamo regime. Figure~\ref{fig:oscillations_Ra5e8_Pm25} shows the time series of the kinetic and magnetic energies of the horizontal wavenumbers $k_h=0$ (for the magnetic field only), $k_h=1$ and $k_h>1$. The initial condition is one of purely hydrodynamic convection, to which a seed magnetic field is added at $t=0$. When the magnetic energy grows to a sufficiently large value, the kinetic energy of the LSV ($k_h=1$) decays rapidly, before reaching a new saturated level, more than two orders of magnitude smaller than in the hydrodynamical case. In the meantime, the kinetic energy of the modes with  $k_h>1$ decreases by about $25\%$, which is mainly attributable to the decrease in the energy of modes $k_h=2-3$. In the saturated phase, the kinetic energy of the mode $k_h=1$ is smaller than the total energy of the other modes by a factor of $30$ (see also the kinetic energy spectra of figure~\ref{fig:specu_Ra5e8}). Although the magnetic field does not completely eliminate the kinetic energy at the largest scale, no coherent LSV is apparent. The magnetic field therefore has an unfavourable effect on the LSV, while leaving the amplitude of the convective flows relatively unchanged.  Since no large-scale magnetic field is generated by this dynamo, the unavoidable conclusion is that the small-scale field is responsible for the suppression of the LSV. It is of interest to note that this suppression does not require a strong magnetic field; at the point when the amplitude of the LSV starts to decrease, the magnetic energy is only about a tenth of the kinetic energy of the convective flow.

\subsection{Oscillations of the LSV dynamo}
\label{sec:5.2}

Here we consider the same convective flow as in \S\,\ref{sec:5.1} (\ie $\Ra=5\times10^8$), but at the much lower magnetic Prandtl number of $\Pm=0.2$.  Time series of the kinetic and magnetic energies are shown in figure~\ref{fig:oscillations_Ra5e8_Pm02}; in contrast to figure~\ref{fig:oscillations_Ra5e8_Pm25}, only the saturated (dynamic) phase is shown, because the kinematic dynamo growth rate is slow. Large oscillations in the kinetic energy of the LSV are observed.  
Such oscillations are absent in the purely hydrodynamic case, and so are indisputably of magnetic origin. 
The kinetic energy of the modes with $k_h>1$ remains mostly unchanged except for small variations that follow
the oscillations of the LSV.
Oscillations of the magnetic energy are associated with the oscillations of the LSV. 
These oscillations are nearly anticorrelated, but a systematic time lag exists between maxima of the LSV amplitude and minima of the magnetic energy.
The oscillations are not quite periodic as
their duration and amplitude 
varies, those with the largest variation having the longest duration. 
All scales of the magnetic field are affected by the oscillations, with a seemingly similar evolution. 
Once the magnetic field is amplified to a critical strength by the dynamo, the LSV decays.  
For each oscillation, a dotted vertical line indicates the approximate time at which the LSV starts to decay.
The vertical lines highlight that this critical value of the magnetic energy is similar for each oscillation ($O(10^{-6})$ for the modes $k_h>1$) 
and is also of the same order as the critical value observed for $\Pm=2.5$.
The decay of the LSV is shortly followed by the decrease of the magnetic energy at all scales.
Subsequently, once the field becomes sufficiently weak, the LSV regenerates, eventually leading to a new phase of growth of the magnetic field. 
While the oscillations of the magnetic energy have sharp minima and maxima, the oscillations of the LSV have
rounded maxima and sharp minima. The rounded maxima are a consequence of the slow growth of the magnetic field, which allows enough time 
during the recovery phase of the LSV for it to saturate hydrodynamically before the magnetic energy reaches the critical value of about $10^{-6}$.
The periodicity of the oscillations is controlled by the growth and decay rates of the magnetic field and also by the growth rate 
of the LSV because the dynamo needs the LSV to have sufficient velocity to start to operate.
In the LSV dynamo --- for which $\Rm$ is below the threshold for small-scale dynamo action --- the small-scale magnetic field is produced by the interactions of the large-scale field with the convective flows. The evolution of the small-scale field therefore closely follows that of the large-scale field. For $\Pm=2.5$, the suppression of the LSV can be attributed solely to the small-scale field; it is plausible that this is also the case for the dynamo at $\Pm=0.2$, although the direct influence of the large-scale field cannot be entirely ruled out. 

\begin{figure}
\centering
  \subfigure[]{\label{fig:oscillations_Ek}
  \includegraphics[clip=true,width=15cm]{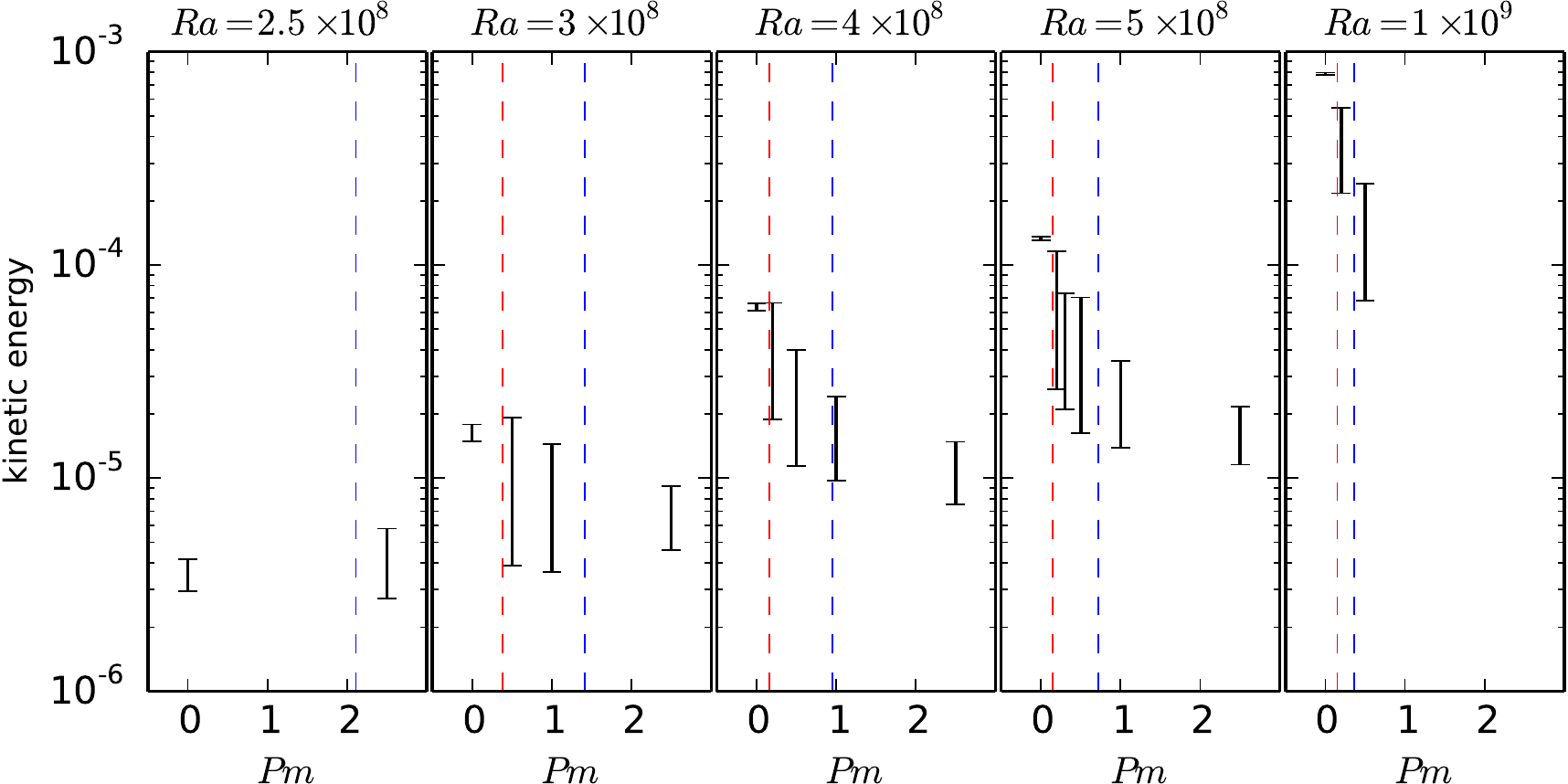}}
  \subfigure[]{\label{fig:oscillations_Em}
  \includegraphics[clip=true,width=15cm]{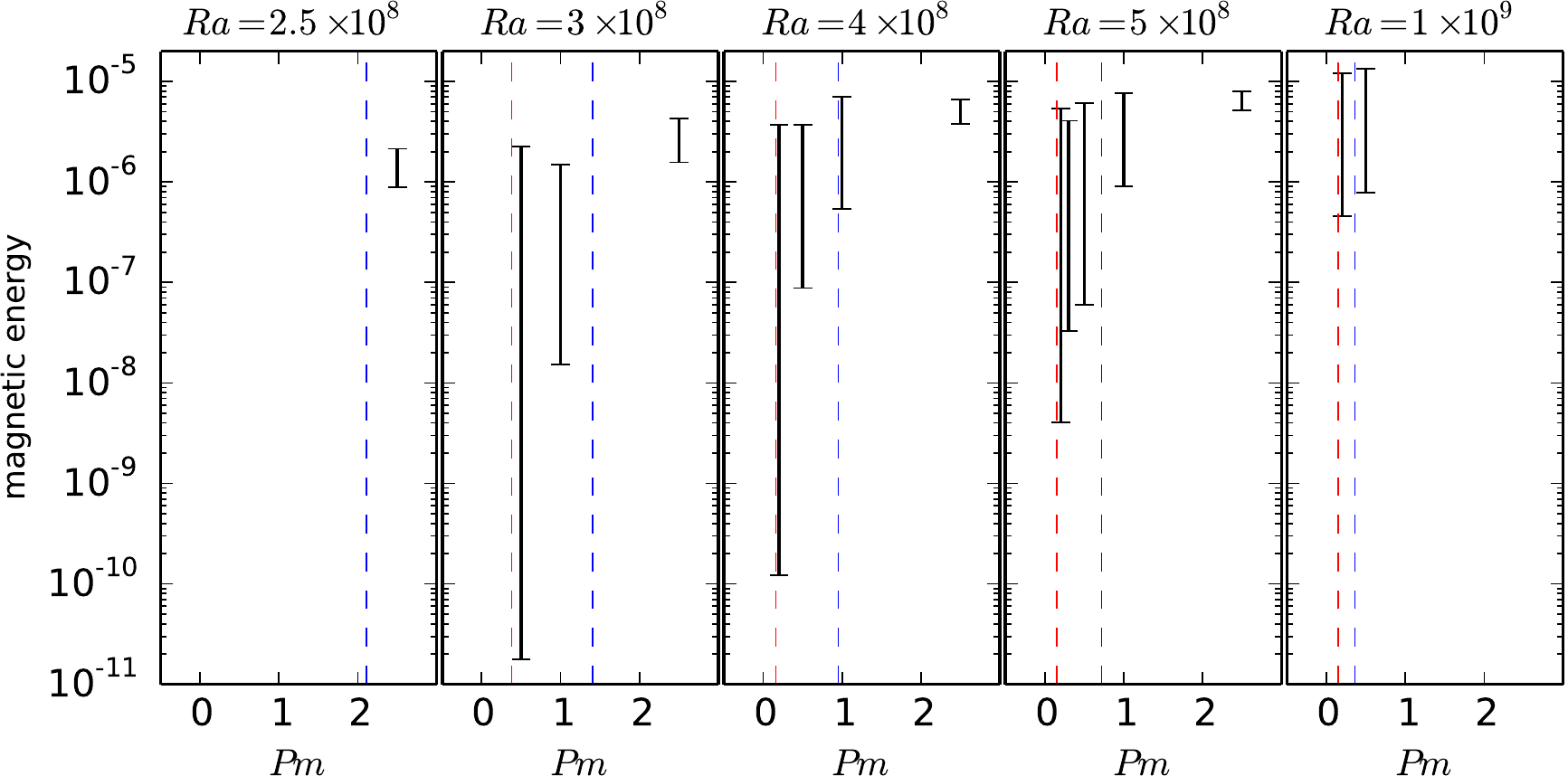}}
  \caption{Range of the oscillations of (a) kinetic and (b) magnetic energy for several values of $\Ra$. The dashed red lines correspond to the threshold of the LSV dynamo and the dashed blue lines to $\Rm=550$, the threshold of the small-scale dynamo (cf. figure~\ref{fig:diagram}).
  For comparison, the relative amplitude of the oscillations of the kinetic energy in the hydrodynamical case is represented at $\Pm=0$.
}
  \label{fig:oscillations_E}
\end{figure}

Figure~\ref{fig:oscillations_E} shows the amplitude of the oscillations of the energies for both the LSV dynamos and small-scale dynamos. 
The transition between the two types of dynamo is identified by the dashed blue line, which corresponds to  $\Rm=550$ (cf.\ figure~\ref{fig:diagram}). 
For comparison, the oscillations of the kinetic energy in the absence of the magnetic field are also represented at $\Pm=0$.
In the absence of the magnetic field, LSVs are present for $\Ra \gtrsim 3 \times 10^8$ and the fluctuations in the kinetic energy are small in this case.
By contrast, the variations of the kinetic energy are much larger in the dynamo cases. In particular, the oscillations of both energies are largest for small $\Pm$. 
Indeed, for small $\Pm$, 
the maximum value of the kinetic energy can even attain the level of the purely hydrodynamic case, implying that the LSV is temporarily fully restored. As $\Pm$ increases, however, the maximum of the kinetic energy remains well below the hydrodynamic level. 
For the LSV dynamos, the maximum of the magnetic energy is comparable at all values of $\Pm$ for a given value of $\Ra$. 
This is consistent with the existence of a critical value of the magnetic field at which the LSV is systematically disrupted. As $\Pm$ increases, this critical value is reached more quickly; as such, the suppression of the LSV occurs before it can reach its full amplitude. Consequently the oscillations tend to have shorter duration as $\Pm$ increases. 

\subsection{Disruption of the Reynolds stresses}
\label{sec:5.3}

In the absence of magnetic field, LSVs are maintained by the persistent action of the Reynolds stresses, and in particular, 
 the interaction of $z$-dependent convective vortices \citep{Rubio14}.
In this subsection, we examine the means by which the magnetic field can lead to the suppression of the LSV by the magnetic field. There are two possible mechanisms: (i) the magnetic field destroys the Reynolds stresses 
that are the source terms of the LSV by decreasing either their amplitude or their spatio-temporal coherence, or (ii) the Maxwell stresses cancel out the Reynolds stresses. To investigate how the Reynolds stresses are modified in the presence of the magnetic field, we focus on two velocity components of the LSV, \mbox{$\langle u_x\rangle_{xz}$} and \mbox{$\langle u_y\rangle_{yz}$}. The equations for the temporal evolution of these components are
\begin{eqnarray}
 	\pdt{\langle u_x\rangle_{xz}} &=& 
	- \frac{\partial}{\partial y} \pleft \langle R - M \rangle_{x} \pright
	+ \Ek  \langle \nabla^2 u_x \rangle_{xz},
 	\label{eq:uxmean}
	\\
 	\pdt{\langle u_y\rangle_{yz}} &=& 
	- \frac{\partial}{\partial x} \pleft \langle R - M \rangle_{y} \pright
	+ \Ek  \langle \nabla^2 u_y \rangle_{yz},
 	\label{eq:uymean}
\end{eqnarray}
where the components of the Reynolds and Maxwell stresses are respectively
\begin{equation}
 	R = \langle u_x u_y\rangle_{z} \quad \textrm{and} \quad M = \langle B_x B_y\rangle_{z},
\end{equation}
and where we have assumed that $\langle u_x\rangle_{yz} \approx 0$ and $\langle u_y\rangle_{xz} \approx 0$. The velocity and magnetic fields are decomposed into $z$-average and $z$-dependent parts, where $v'=v-\langle v\rangle_z$. The stress components $R$ and $M$ can then be decomposed into
\begin{eqnarray}
 	R &=& R_m + R_f, \quad \textrm{with} \quad 
	R_m = \langle u_x \rangle_{z}  \langle u_y \rangle_{z}, \quad
	R_f= \langle u'_x u'_y \rangle_{z},
	\label{eq:R}
	\\
	 M &=& M_m + M_f, \quad \textrm{with} \quad 
	M_m = \langle B_x \rangle_{z}  \langle B_y \rangle_{z}, \quad
	M_f= \langle B'_x B'_y \rangle_{z}.
	\label{eq:M}
\end{eqnarray}

\begin{figure}
\centering
  \includegraphics[clip=true,width=13cm]{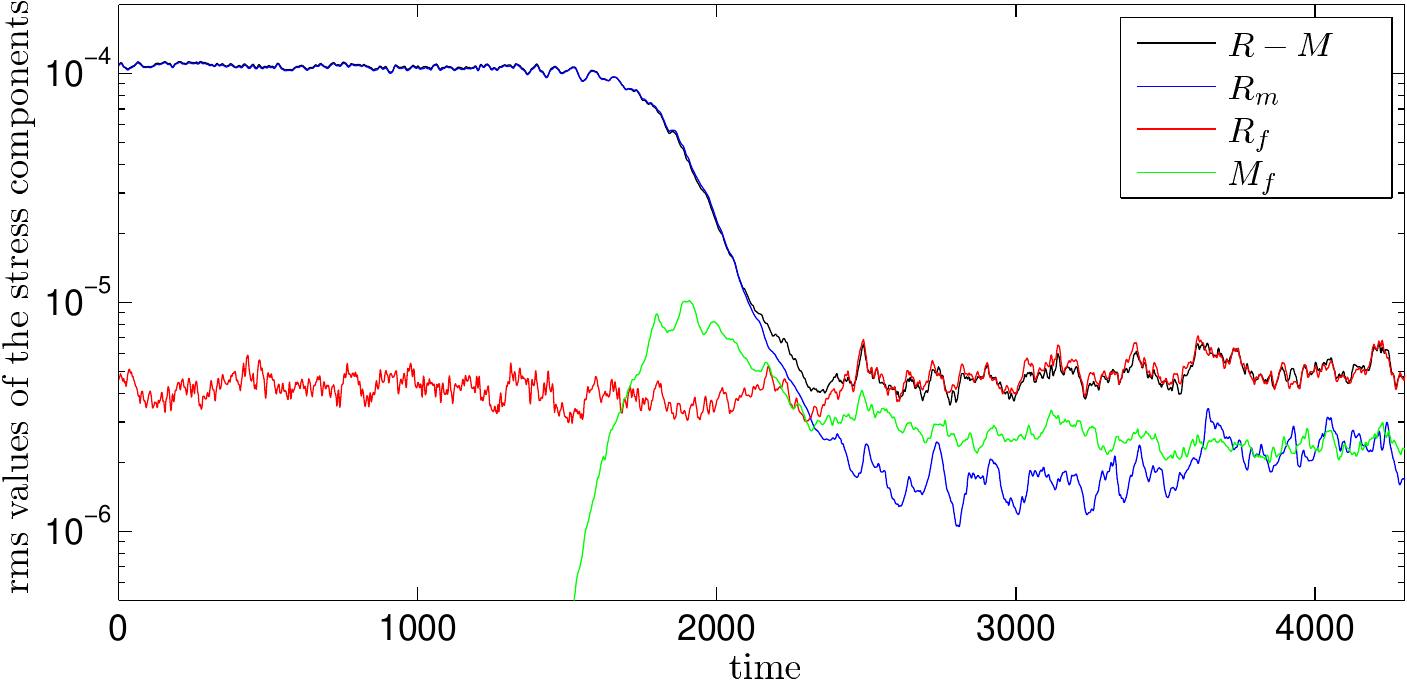}
  \caption{Time series of the rms values of the components of the Reynolds and Maxwell stresses as defined in
  equations (\ref{eq:R}) and (\ref{eq:M}) for the same time series as in figure~\ref{fig:oscillations_Ra5e8_Pm25}.}
  \label{fig:stress}
\end{figure}

The term of particular interest is $R_f$, which represents the interaction of $z$-dependent vortices, and thus, the driving term of the LSV.
Figure~\ref{fig:stress} shows the time evolution of the rms values of $R_m$, $R_f$ and $M_f$ 
for the small-scale dynamo at $\Ra=5\times10^8$ and $\Pm=2.5$. This time series can be compared with the evolution of the energies in figure~\ref{fig:oscillations_Ra5e8_Pm25}. Since the magnetic field does not have a large-scale component, the rms value of $M_m$ is small, and hence is not plotted in figure~\ref{fig:stress}. 
As the small-scale field grows, the most notable change to the Reynolds stresses is the decrease of the rms value of $R_m$ by about two orders of magnitude.
This is an expected consequence of the decrease of the velocity of the LSV by an order of magnitude since the major contribution to $R_m$ comes from the self-interaction of the LSV.
Of greater interest are the more subtle changes to the rms value of $R_f$:
the amplitude of $R_f$ remains at 
a similar level in the presence of the magnetic field but its temporal variations change, marked in particular by the disappearance of the high frequencies. 
The suppression of the LSV cannot therefore be ascribed to a decrease of the amplitude of its driving term $R_f$.
As the magnetic field increases in strength, $M_f$ increases, eventually reaching a level about half the averaged rms value of $R_f$.
The amplitude of $M_f$ is therefore
too small to cancel completely $R_f$, but partial cancellation might occur. 
To determine this, the rms value of the residual of the components of the Reynolds and Maxwell stresses, $R-M$, is also plotted in figure~\ref{fig:stress}.
$R-M$ has the same amplitude as $R_f$ in the saturated phase of the dynamo, thereby indicating that the suppression of the LSV is not due to a straightforward cancellation of the Reynolds stresses by the Maxwell stresses. 
Instead, it follows that the LSV is suppressed by a modification of the driving term $R_f$ by the magnetic field, although its amplitude remains relatively unchanged.

\begin{figure}
\centering
  \subfigure[]{
  \includegraphics[clip=true,width=\textwidth]{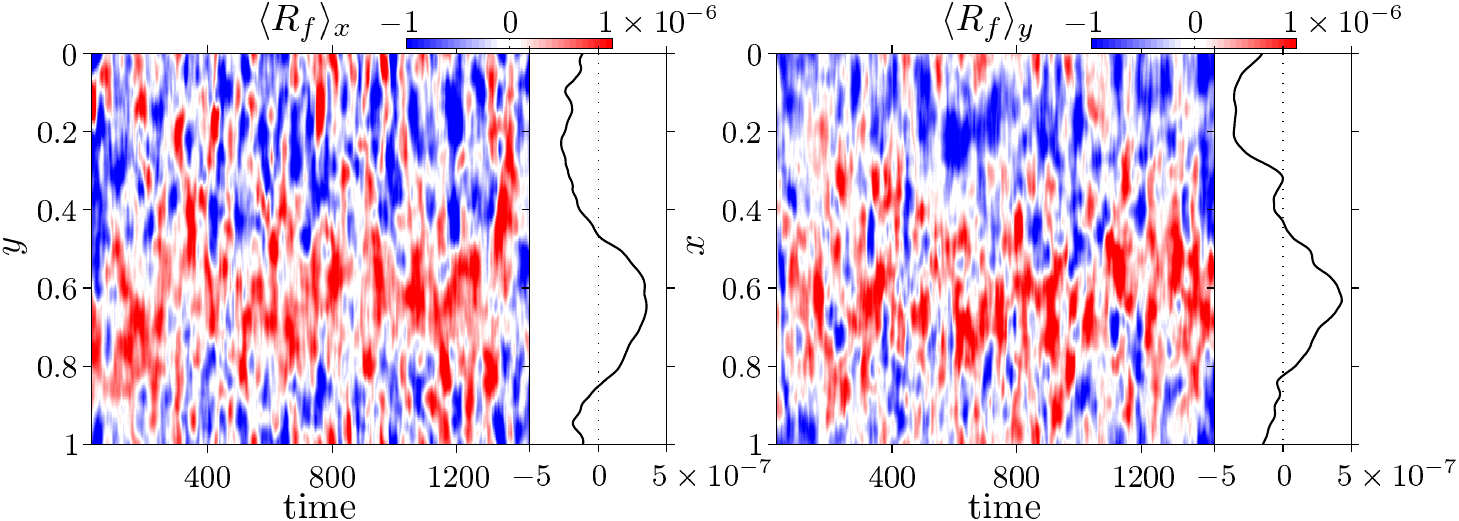}}
  \subfigure[]{
  \includegraphics[clip=true,width=\textwidth]{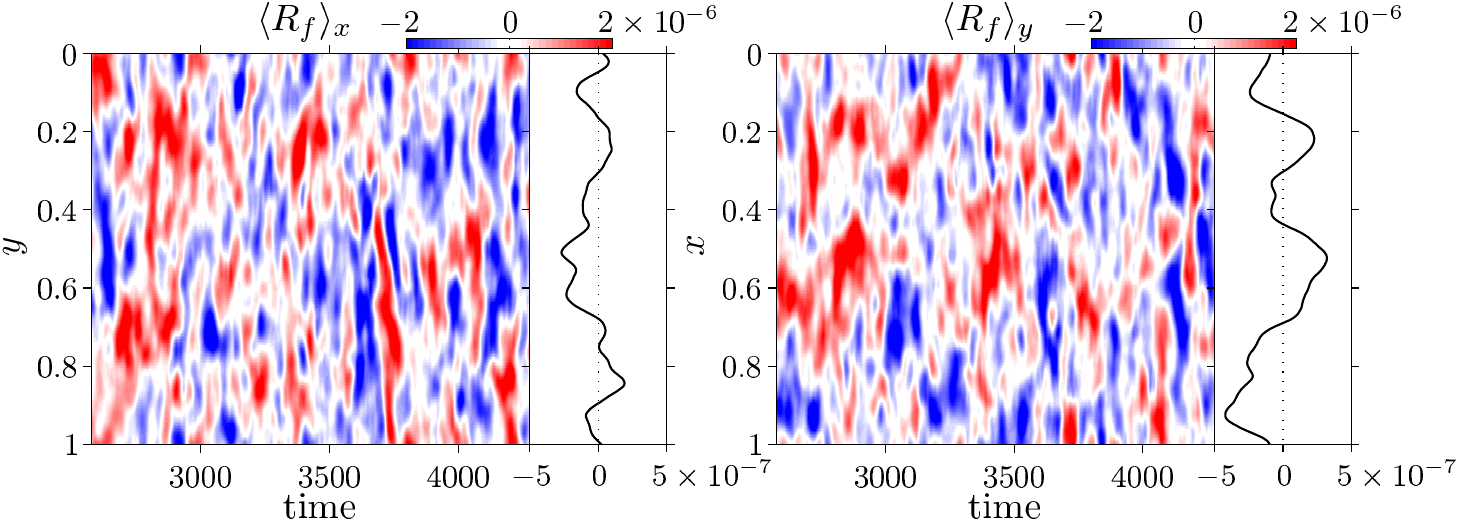}}
  \caption{Space-time diagram of $\langle R_f \rangle_x$ and $\langle R_f \rangle_y$  for the same series 
  as figure~\ref{fig:stress}: (a) the kinematic phase, and (b) the saturated phase of the dynamo. The plot on the right of each diagram represents the time average.}
  \label{fig:Rxy_time}
\end{figure}

Figure~\ref{fig:Rxy_time} shows the spatio-temporal evolution of $\langle R_f\rangle_x$ and $\langle R_f \rangle_y$ during the kinematic and saturated phases of the dynamo. To remove the variations of high temporal frequency in the data, we have performed a moving average over $30$~time units. During the kinematic phase, $\langle R_f\rangle_x$ and $\langle R_f\rangle_y$ maintain good temporal coherence of the modes $k_y=1$ and $k_x=1$ respectively. The time series is short enough that the LSV does not drift appreciably in the horizontal plane. The time averages therefore have a well-defined signal at $k_y=1$ or $k_x=1$. Thus, in the kinematic regime, the LSV is forced consistently by the convective vortices. By contrast, during the saturated phase, this temporal coherence is lost, so the LSV cannot be forced consistently. Thus, significantly, the demise of the LSV is due not to a reduction in the amplitude of the Reynolds stresses that feed the LSV, nor by their cancellation by the Maxwell stresses, but instead to the more subtle influence of the small-scale field in hindering the coherence of the Reynolds stresses.
We note that the study of \citet{Rubio14} in non-magnetic convection shows that the feedback of the LSV itself plays a positive role on the energy 
transfer from small to large scales. 
The loss of the positive feedback of the LSV might therefore be partly responsible for the loss of the coherence in the Reynolds stresses,
although its primary cause is the disruption by the magnetic field.

\section{Discussion}
\label{sec:ccl}

In \citet{Guervilly2014}, we showed how large-scale vortices (LSVs) can be spontaneously generated from a hydrodynamical process that consists of the clustering of small-scale, rotationally constrained convective vortices. In this paper, we have explored the dynamo action that may result from turbulent rotating convection in the presence of LSVs. For a range of magnetic Reynolds numbers above the critical value for the onset of dynamo action ($100 \lesssim \Rm \lesssim 550$ for $\Ek=5\times10^{-6}$ and $\Pran=1$), the flow acts to generate a magnetic field with a significant component on scales large compared with the small-scale convective vortices --- we denote this as an LSV dynamo. 
The dynamo generates magnetic fields that are concentrated in the shear layers surrounding the LSV, together with a coherent mean (\ie horizontally-averaged) magnetic field. From considering the kinematic dynamo problem with spectrally filtered versions of the velocity, we find that the dynamo mechanism requires only the presence of the LSV together with velocity modes of scale intermediate between the box size and the dominant convective scale. When the LSV is artificially filtered out from the induction equation but the effect of the LSV on the flow is retained in the momentum equation, the dynamo fails. Consequently, the LSV plays a crucial role in the magnetic induction; importantly, this is not simply via its action on the smaller-scale flows.
The filtered simulations indicate that the magnetic field is generated in the core of the LSV and in the surrounding shear layers, having been expelled from the core by small-scale vortices.
These results are deduced from the filtering exercise performed in the kinematic phase of the dynamo, so they could be specific to this phase. However, the qualitative similarities 
of the magnetic field in the filtered kinematic dynamo simulations and in the full dynamic regime suggest that the dynamo mechanism operates in a similar manner in both cases.

For $\Rm \gtrsim 550$, the convective flows generate a small-scale dynamo. In this case, the continuous production of the small-scale magnetic field acts to suppress the LSV completely. LSVs are produced by the Reynolds stresses resulting from the interactions of depth-dependent vortices; the small-scale magnetic field hinders these interactions, leading to a loss of coherence of the Reynolds stresses, and hence an inability to create an LSV. The LSV dynamo can therefore operate only below the threshold for small-scale dynamo action. The transition from the LSV dynamo to the small-scale dynamo is continuous. In the LSV dynamo, oscillations of the kinetic and magnetic energies are associated with 
cycles of suppression and regeneration of the LSV. These oscillations are of magnetic origin, and are due to the amplification of the small-scale magnetic field from the interactions between the large-scale field and the convective flows. 
The suppression of the LSV by small-scale magnetic fields at high $\Rm$ therefore probably limits the relevance of the LSV dynamo mechanism to astrophysical objects with moderate $\Rm$. Such is the case of planetary dynamos, and in this context, the ability of the LSV dynamo to operate at low magnetic Prandtl numbers is of great interest. 
Although the values of $\Pm$ in our simulations are a long way from realistic values of the order of $10^{-6}$,
it is entirely possible that, at smaller $\Ek$, LSV dynamos might be found at lower $\Pm$.

In the Cartesian geometry employed here, the LSV dynamo produces a mean horizontal magnetic field with a coherent `staircase' structure that overall rotates clockwise. The spatial and temporal variations of this mean field are more complex than those of the mean field generated by the large-scale dynamo that operates near the onset of convection (compare, for instance, figure~\ref{fig:LSdynamo_c} and figure~\ref{fig:LSVD_Bmean_b}). This latter dynamo relies on the coordinated action of the convective vortices, and so can operate only for slightly supercritical convection (up to 50\% above the critical Rayleigh number for the parameters considered here). Its mean field always has a simple spatial structure with a regular time dependence, quite unlike the spatial and temporal variations of planetary magnetic fields. From this point of view, the greater complexity of the mean magnetic field of the LSV dynamo is thus an interesting feature, although the Cartesian geometry is inadequate for a comparison with planetary magnetic fields.

The choice of boundary conditions may have an influence on the formation of LSVs and any subsequent dynamo action. The role of the velocity boundary conditions in LSV formation has been addressed in the recent paper by \citet{Stellmach2014}.
The magnetic boundary conditions used here enforce a purely horizontal magnetic field at the top and bottom boundaries and these 
boundary conditions might have an effect on the LSV dynamo. To examine the role of the magnetic boundary conditions, we have also performed a simulation with boundary conditions that enforce
a purely vertical magnetic field at the top and bottom boundaries ($B_x=B_y=\partial_z B_z=0$ at $z=0,1$) for the same parameters as the dynamo of figure~\ref{fig:LSVD_Bmean} 
($\Ra=5\times10^8$ and $\Pm=0.2$).  We find that the generated magnetic field retains its main characteristics: a similar saturation level of the magnetic energy, 
concentration of the magnetic field in the shear layers around the LSV and a coherent mean horizontal magnetic field. 
The main differences are that the mean horizontal field tends to be symmetric with respect 
to the mid-plane and that a stronger vertical magnetic field is produced, with an average energy about half that of the horizontal field. Thus, overall, the magnetic boundary conditions (at least the two types tested here) have only a minor influence on the LSV dynamo mechanism.

The choice of the aspect ratio of the numerical domain might also influence the LSV dynamo, since the width and strength of the LSV both increase 
with aspect ratio \citep{Guervilly2014}. To explore this question, however,  requires a series of lengthy calculations, going beyond the scope of the present study; 
it is though an issue that we propose to examine in future work.

One of the most interesting aspects of the recent work on LSV formation in rotating convection, together with the resulting dynamos considered in this paper, is the discovery of new physical phenomena at the small values of the Ekman number that can now be tackled numerically. It is therefore instructive to consider how the formation of LSVs and the ensuing dynamos might be affected at yet lower values of $\Ek$ --- though clearly one cannot rule out the appearance of yet further novel behaviour in the considerable gap that exists between what is currently computationally feasible ($\Ek =O(10^{-6})$) and what is appropriate for the Earth ($\Ek =O(10^{-15})$), for example. 
In terms of the LSVs, it is clear that they require vigorous, yet rotationally constrained turbulence; as such we can be confident that the range of Rayleigh numbers at which LSVs will be found will increase as $\Ek$ is decreased, as portrayed in figure~\ref{fig:diagram_LSV}. The issue of the LSV dynamo is, however, less clear-cut. Although we may expect the LSV dynamo to operate for $\Ra$ just above the value at which LSV can be formed, it is less straightforward to predict the onset of small-scale dynamo action, and hence the demise of the LSV dynamo.

The set-up considered here --- i.e., a Boussinesq fluid in Cartesian geometry --- forms the simplest system in which to study dynamos driven by rotating convection; it thus allows for a relatively wide exploration of parameter space. However, spherical geometry is clearly more appropriate for the study of planetary cores. To date, LSVs have not been reported in numerical simulations of rotating convection in spherical geometry; large-scale coherent flows \textit{are} observed for low $\Ek$ and large $\Rey$, but these are zonal flows --- \ie axisymmetric and azimuthal jets \citep[e.g.][]{Heimpel05, Gastine2012}. 
Interestingly, these zonal flows are also known to be disrupted by magnetic fields in spherical convective dynamos \citep[e.g.][]{Aub05,Yad16}.
In terms of rotating convection, one of the major differences between Cartesian and spherical geometries is the anisotropy caused by spherical geometry in the plane normal to the rotation axis. This anisotropy constrains the geostrophic flows to be azimuthal, unlike in Cartesian geometry, where there is no preferred horizontal direction. Consequently, zonal flows tend to be dominant in rotating spherical convection, provided that the viscous damping from the boundary layers is not too large, \ie for stress-free boundary conditions. Interestingly, however, models of barotropic flows on a $\beta$-plane with forced stirring have shown that non-zonal coherent structures of low azimuthal wavenumber can co-exist with dominant zonal jets \citep{Galerpin10, Bakas14, Constantinou15}. These large-scale coherent structures take the form of propagating waves, which are believed to be sustained by nonlinear interactions between Rossby waves. One may thus speculate that at sufficiently small values of $\Ek$, large-scale vortices, and the resultant dynamos, may indeed play a role in spherical geometry; if so, we might expect them to have a propagating  nature, a feature that cannot be  recovered in plane parallel geometry.

\section*{Acknowledgements}
This work was supported by the Natural Environment Research Council under grant NE/J007080/1 and NE/M017893/1. This work was undertaken on ARC1 and ARC2 (part of the High Performance Computing facilities at the University of Leeds), the UK National Supercomputing Service ARCHER, and COSMA Data Centric system. COSMA is operated by the Institute for Computational Cosmology at Durham University on behalf of the STFC DiRAC HPC Facility (www.dirac.ac.uk). This equipment was funded by a BIS National E-infrastructure capital grant ST/K00042X/1, DiRAC Operations grant ST/K003267/1 and Durham University. DiRAC is part of the National E-Infrastructure.
We are grateful to the referees for helpful comments that improved the presentation of the paper.

\end{document}